\documentclass[reprint,amsmath,amssymb,aps,pre,longbibliography]{revtex4-1}

\usepackage{graphicx}
\usepackage{dcolumn}
\usepackage{bm}

\newcommand{\p}[1]{\partial_{#1}}
\newcommand{\Ro}{\text{Ro}}

\begin{document}
\preprint{APS/123-QED}
\title{Statistics of inhomogeneous turbulence in large scale quasi-geostrophic dynamics}
\author{Anton Svirsky$^{1}$, Corentin Herbert$^{2}$ and Anna Frishman$^1$}
\email{frishman@technion.ac.il}

\affiliation{$^1$Physics Department, Technion Israel Institute of Technology, 32000 Haifa, Israel}
\affiliation{$^2$Univ Lyon, ENS de Lyon, CNRS, Laboratoire de Physique, F-69342 Lyon, France}

\date{\today} 

\begin{abstract}
A remarkable feature of two-dimensional turbulence is the transfer of energy from small to large scales. This process can result in the self-organization of the flow into large, coherent structures due to energy condensation at the largest scales. We investigate the formation of this condensate in a quasi-geostropic flow in the limit of small Rossby deformation radius, namely the large scale quasi-geostrophic model. In this model potential energy is transferred up-scale while kinetic energy is transferred down-scale in a direct cascade. We focus on a jet mean flow and carry out a thorough investigation of the second order statistics for this flow, combining a quasi-linear analytical approach with direct numerical simulations. We show that the quasi-linear approach applies in regions where jets are strong and is able to capture all second order correlators in that region, including those related to the kinetic energy. This is a consequence of the blocking of the direct cascade by the mean flow in jet regions, suppressing fluctuation-fluctuation interactions. The suppression of the direct cascade is demonstrated using a local coarse-graining approach allowing to measure space dependent inter-scale kinetic energy fluxes, which we show are concentrated in between jets in our simulations. We comment on the possibility of a similar direct cascade arrest in other two-dimensional flows, arguing that it is a special feature of flows in which the fluid element interactions are local in space.
\end{abstract}
\maketitle

\section{Introduction}
Multi-scale, non-linear, interactions are one of the defining properties of turbulent flows, posing a considerable challenge both for theoretical understanding and numerical modeling.
In particular, they imply that the flow at large scales is coupled to smaller scale fluctuations, and that, for instance, the structure of the large scale flow depends on the transport of momentum and the dissipation of energy by such small scales.
For statistically homogeneous and isotropic flows such interactions are, to leading order, well captured within phenomenological theories, though many theoretical questions remain open and there are very few results which can be obtained from first principles~\cite[e.g.]{Frisch1995, pope_turbulent_2000}. 
However, most real flows break such symmetries at large enough scales, either because of external fields such as gravity or a magnetic field, due to the effect of rotation, or the existence of boundaries.
In such flows, often the key question is to characterize the large scale mean-flow. In turn, this requires the prediction of energy and momentum transfers due to turbulent fluctuations, across scales and spatially. Generally, this is a challenging task, requiring ad-hoc assumptions. However, there is growing evidence that the non-linear interactions in the presence of a strong mean flow may in fact be easier to treat than those in a homogeneous and isotropic flow (see e.g.~\cite{marston_recent_2023} for a review). This has been particularly evident 
in two-dimensional (2D) and quasi-2D flows, where dimensionality imposes strong constraints upon the nature of multi-scale interactions.

Two-dimensional flows exhibit a remarkable tendency to spontaneously self-organize into a coherent mean flow when excited at small scales. The mechanism behind this self-organization is an inverse transfer of a quadratic invariant (e.g energy) from small to large scales, in a process called the inverse cascade~\cite{kraichnan_inertial_1967,Leith1967,Batchelor1969}. The inverse cascade arises due to the existence of a second inviscid invariant of the dynamics, which is simultaneously transferred to small scales in a so-called direct cascade. In a finite system this inverse transfer results in the accumulation of energy at the largest available length scale, forming a system-size coherent mean flow termed a condensate~\cite{kraichnan_inertial_1967, bose_smith_1993, dynamics_chertkov_2007,Chertkov2010}.
In this condensate regime, the direct interactions between the mean-flow and turbulence can dominate over local-in-scale interactions.
Indeed, theoretical ideas and numerical methods which do not explicitly resolve the fluctuation-fluctuation interactions have been shown to be applicable in this system (and its variants)~\cite{Farrell2003,Farrell2007,Tobias2013,Marston2016,Srinivasan2012,Bouchet2013}.
Moreover, analytical results describing both the spatial structure of the mean flow~\cite{laurie_universal_2014, kolokolov_structure_2016, frishman_culmination_2017} and the turbulent kinetic energy density~\cite{frishman_turbulence_2018}, were successfully derived from first-principles in this regime.

The interest in two-dimensional flows is not limited to the theoretical understanding of turbulent interactions, 
as many flows in nature become effectively two-dimensional and thus exhibit a similar phenomenology. This occurs when the fluid motion is constrained in one of the directions either because the fluid is contained within a thin layer, is stratified, or is rapidly rotating~\cite{xia_spectrally_2009,boffetta_two_2012,rivera_direct_2014,bouillaut_experimental_2021}. Astrophysical and large-scale geophysical flows often have one or more of these properties, with rotation playing a particularly important role in constraining the motion, called the geostrophic regime~\cite{young_forward_2017,vallis_atmospheric_2017}.
A minimal model for the flow in this regime, capturing the main features of the large-scale dynamics and serving as an important theoretical tool, is the shallow water quasi-geostrophic equation (SWQG).
In SWQG, there is a typical scale which determines the range of interactions between fluid elements, called the Rossby deformation radius.
The special cases within which the condensate had been studied in detail so far are two-dimensional incompressible Navier-Stokes (2DNS) with the beta effect (differential rotation)~\cite{Farrell2007,Srinivasan2012,Tobias2013,Woillez2017, Woillez2019}, or 2DNS without differential rotation~\cite{laurie_universal_2014,frishman_turbulence_2018,frishman_culmination_2017}.
Both these cases capture dynamics at scales much smaller than the deformation radius, so that interactions span the entire domain.

Here we consider the opposite limit, where interactions are local--- the so-called large-scale quasi-geostrophic (LQG) equation~\cite{larichev_weakly_1991}. This model captures the long-time dynamics at scales much larger than the deformation radius. It contains two inviscidly conserved quantities: the potential energy, transferred to large scales, and the kinetic energy, which cascades to small scales in a direct cascade.
The main question we investigate is what type of condensate does this system support, and how does the locality of interactions affect the properties of turbulent fluctuations and their interaction with the mean-flow.

We begin by reviewing the derivation of the LQG equation from SWQG and discuss the conditions necessary for the emergence of a condensate in this system. Building upon our results for the mean flow of an LQG jet-type condensate~\cite{SI}, we fully characterize the two-point second-order statistics, combining analytical derivations and results from direct numerical simulations.
Turning to the direct cascade and the kinetic energy balance, we show that in the presence of a mean flow it includes a spatial flux of fluctuating kinetic energy. Such a flux is absent in 2D Navier-Stokes, and seems to be related to the locality of interactions in LQG. Using our analytical results for the second-order statistics, we show that this flux carries most of the kinetic energy away from regions of strong mean flow, effectively arresting the direct cascade there. We confirm the in-homogeneity of the direct cascade induced by the mean flow by examining the flux of energy between scales within a smooth filtering approach~\cite{eyink_localness_2009}. In particular, we consider the local inter-scale flux of kinetic energy and potential energy for the LQG system. Measuring this flux in simulations, we demonstrate that the flux of kinetic energy to small scales is indeed locally suppressed in regions where the jets are strong, an effect so strong it is evident under short-time averaging.

This work also serves as the companion to the paper~\cite{SI}. Here we provide detailed derivations and discussions of some of the results stated in~\cite{SI} alongside new results and analysis not contained in~\cite{SI}.

\section{Framework}
The LQG equation can be derived as a limit of the shallow water quasi-geostrophic equation which is given by~\cite{vallis_atmospheric_2017}
\begin{equation}
    \partial_t q+\bm{v}\cdot \nabla q=\partial_{t}q+J(\psi,q)=0; \quad q=\left(\nabla^{2}-L_{d}^{-2}\right)\psi,
\end{equation}
where $q$ is the potential vorticity, $\psi$ is the stream-fucntion which is related to the fluid velocity via $\bm{v} = \bm{\hat{z} \times \bm{\nabla}}\psi$, $\omega=\nabla^{2}\psi=\left(\boldsymbol{\nabla}\times\boldsymbol{v}\right)\cdot\boldsymbol{\hat{z}}$ is the vorticity, $J(\psi,q)$ is the Jacobian operator defined as $J(\psi,q)=\partial_{x}\psi\partial_{y}q-\partial_{y}\psi\partial_{x}q=\epsilon_{ij} \partial_i \psi \partial_j q$ with $\epsilon_{ij}$ the 2D Levi-Civita symbol.
This equation describes the dynamics of a rapidly-rotating homogeneous fluid layer, wherein the pressure gradient force due to fluctuations of the free surface is balanced by the Coriolis force (the so-called \emph{geostrophic balance}).
The stream function both determines the velocity and is proportional to the surface height perturbations of the fluid layer. The scale $L_d$ is called the Rossby deformation radius and sets the range of influence of a surface perturbation on its surroundings. When $L_{d}/L\to\infty$, where $L$ is a characteristic scale for the domain size,  surface perturbations have a long-range influence on the fluid, and their equilibration is fast compared to the rotation period, giving an incompressible 2D fluid. The opposite limit $L_{d}/L\to0$, corresponds to a very rapidly rotating fluid, where the effect of surface perturbations is strictly local. In this limit, the long time dynamics are given by the LQG equation~\cite{larichev_weakly_1991}
\begin{equation}
\label{eq:LQG}
\partial_\tau\psi + \bm{v}^\omega \cdot \bm{\nabla} \psi=\partial_{\tau}\psi+J(\omega,\psi)=f +\alpha\nabla^{2}\psi-\nu ( -\nabla^{2})^{p}\psi,
\end{equation}
with $\bm{v}^\omega = \bm{\hat{z} \times \bm{\nabla}}\omega$ and we have included forcing $f$ and dissipation, with $\alpha$ the friction  coefficient (corresponding to linear drag on velocity $\bm{v}$) and $\nu$ the (hyper) viscosity.

This advection equation is similar to 2DNS but with the roles of the vorticity and the stream-function reversed. Here the vorticity acts as the "effective stream function", and the stream-function is advected by an effective velocity $\bm{v}^\omega = \bm{\hat{z} \times \bm{\nabla}}\omega$. The integral invariants of \eqref{eq:LQG} without forcing and dissipation are the kinetic energy $Z=\frac{1}{2}\int\left(\nabla\psi\right)^{2}\text{d}^{2}x=\frac{1}{2}\int|\bm{v}|^{2}\text{d}^{2}x$ and all moments of $\psi$, in particular the potential energy $E=\frac{1}{2}\int\psi^{2}\text{d}^{2}x$. The existence of the two quadratic invariants results in the inverse cascade of $E$ and a direct cascade of $Z$ \cite{smith_turbulent_2002}.

Let us briefly comment on the conditions for the LQG limiting dynamics to be consistent, a more complete discussion can be found in the Appendix~\ref{appendix: LQG_from_SWQG}. The SWQG equation is derived from the rotating shallow water equations in the limit of a small Rossby number, $\text{Ro}=U/(\Omega L)$ with $U$ a typical velocity scale, and $\Omega$ the fluid rotation rate. The derivation also requires that $\text{Ro} (L/L_d)^2\sim o(1) $ so that height perturbations are small compared to the mean fluid thickness. The limit $L_d/L\to 0$ which we take to obtain LQG is consistent with this assumption provided that $L_d/L\sim \text{Ro}^\beta$ with $0<\beta<1/2$, under which condition LQG can be derived as a limit of SWQG . We remark that traditionally SWQG is derived assuming $L_d / L \sim O(1)$ rather than $L_d/L\sim \text{Ro}^\beta$. However, we show in the Appendix~\ref{appendix: LQG_from_SWQG} that LQG can be directly derived from the rotating shallow water equations in the latter limit, that requires rescaling time by $\tau = t(L_d/L)^2\propto t\text{Ro}^{2\beta}$ and expanding the height field in powers of $\text{Ro}^{n+2\beta}$ (instead of $\text{Ro}^{n}$ like the velocity).
Thus, the forced LQG equation should be able to capture the large scale dynamics of SWQG with a small but finite $L_d$ with a forcing scale which is larger than $L_d$~\cite{smith_turbulent_2002}. Note that there is evidence that the inviscid equation eventually develops motions on smaller scales, so the LQG equation may become inadequate~\cite{burgess_potential_2022}.

We wish to explore the LQG system \eqref{eq:LQG} in the condensation regime, where the potential energy condenses at the largest available scale. This requires that the rate of energy removal at the box scale is much slower than the transfer rate by the inverse cascade. The inverse cascade rate (eddy-turnover time) can be found using dimensional analysis, requiring that this rate depends only on the scale and the potential energy injection rate $\epsilon = \langle \psi f \rangle$ ~\cite{smith_turbulent_2002}. We have $[\epsilon]\sim[\psi^2]/t\sim l^8/t^3$, so that the rate of the inverse cascade of $E$ at scale $l$ is $\tau_E(l) \sim \epsilon^{-1/3}l^{8/3}$. Similarly, the rate of the direct cascade of $Z$ at scale $l$ is $\tau_Z(l) \sim \eta^{-1/3}l^2$, where $\eta = \langle \bm{\nabla}\psi \bm{\nabla}f \rangle$ is the kinetic energy injection rate. Assuming the forcing acts in a narrow band of scales around $l_f$, the injection rates can be simply related by $\eta \approx \epsilon/l_f^2$. Note that the eddy turnover time decreases with the scale much faster than in 2DNS, with a factor of $l^2$ between the two. This limits the available resolution for simulations and thus also the separation of scales between the forcing and box scale.

The dissipation rates due to the drag and viscous terms are  $\tau_\alpha(l) \sim \alpha^{-1} l^2$ and $\tau_\nu(l) \sim \nu^{-1} l^{2p}$ respectively. Assuming $p\geq 2$ (integer), the former will serve as the large scale dissipation mechanism arresting the inverse cascade of $E$ while the latter as the small-scale dissipation arresting the direct cascade of $Z$. For the potential energy to condense at the box scale, $L$, requires  $ \delta \equiv \tau_E(L)/\tau_\alpha(L) = \alpha(L^2/\epsilon)^{1/3} \ll1$. Note that $\delta$ grows with the scale, i.e. that the ratio between the non-linear time-scale and the dissipative time scale grows with the length scale (like in 2DNS with linear friction, factors of $l^2$ appearing in both time-scales cancelling out to give the same ratio). Additionally, in order for a significant fraction of the (potential) energy to be transferred to large scales, we require that at the forcing scale the dissipation rate $\tau_\nu(l_f)$ is low compared to the non-linear transfer rate, resulting in the requirement $\text{Re} \equiv \tau_\nu(l_f)/\tau_E(l_f) =  l_f^{2p-8/3}\epsilon^{1/3}/\nu \gg 1$.
The kinetic energy cascade is arrested at the Kolmogorov scale $l_\nu$ where the inverse cascade rate and the viscous dissipation rate are comparable, resulting in  $l_{\nu} \sim (\nu^3/\epsilon)^{1/(6p-8)}$. We can then estimate the potential energy dissipation at small scales: assuming a constant kinetic energy flux down to the Kolmogorov scale, the energy dissipation rate is given by $\epsilon_{\nu}=l_{\nu}^2\eta$ where $\eta=\epsilon/l_f^2$ is the injected kinetic energy. Thus we get $\epsilon_{\nu}=(l_{\nu}/l_f)^2 \epsilon$. This implies that the ratio between the dissipated energy and the energy transferred to large scales is $\epsilon_\nu/\epsilon_\alpha=\epsilon_\nu/(\epsilon-\epsilon_\nu)=l_\nu^2/(l_f^2-l_\nu^2)\approx l_\nu^2/l_f^2$ which is indeed small in the limit of a large Re number. The sharpness of the small-scale cutoff is determined by $p$, higher values will increase the kinetic energy removal rate at scales $l<l_{\nu}$ (and decrease the removal rate for $l>l_{\nu}$) resulting in a sharper cutoff of the spectrum at $l_{\nu}$.

We perform direct numerical simulations (DNS) of the LQG equation \eqref{eq:LQG} using the Dedalus framework \cite{burns_dedalus_2020}. The pseudo-spectral method is implemented using the 3/2 dealiasing rule and time stepping using a third-order, four-stage DIRK/ERK method.
We focus on a jet-type LQG condensate, which simplifies the analysis in the following. In a doubly periodic domain such a condensate emerges if the symmetry between the $x$ and $y$ directions is broken~\cite{bouchet_random_2009,frishman_jets_2017}.
Note that such jets differ from those which emerge due to differential rotation (beta-plane turbulence~\cite{Rhines1975}), which is absent here.
We therefore use a doubly periodic box of dimensions $L\equiv L_{y}=2L_x=2\pi$. The spatial resolution is taken to be $64\times128$, which is relatively low, restricted by the rapid decrease of the eddy-turnover time with decreasing scale in LQG. We use a white-in-time forcing which is localized in Fourier space at a wavenumber $k_{f} =2\pi/l_f = (10,13,15)$ (forcing in an annulus of width $2dk=2$ with a constant amplitude $A=10^{-3}$ and a random phase). We use hyper-viscosity with $p=7$ and $\nu =(0.5,7.3,10)\times10^{-19}$ and take $\alpha=(0.5,1,2)\times 10^{-3}$. Simulation parameters are chosen such that a significant fraction of the (potential) energy is transferred to large scales $Re\gg 1$ and such that potential energy condenses at large scales $\delta  \ll1$.
 Each simulation is run until the system reaches a statistically steady state, and statistics are then gathered over many large scale turnover-times $\tau_E(L)$. The full list of simulations performed is presented in Table~\ref{tab:sim-list} and the choice of the temporal and spatial resolutions are discussed in Appendix~\ref{appendix: resolution}.

\begin{figure}[b t!]
\includegraphics[width=1\columnwidth]{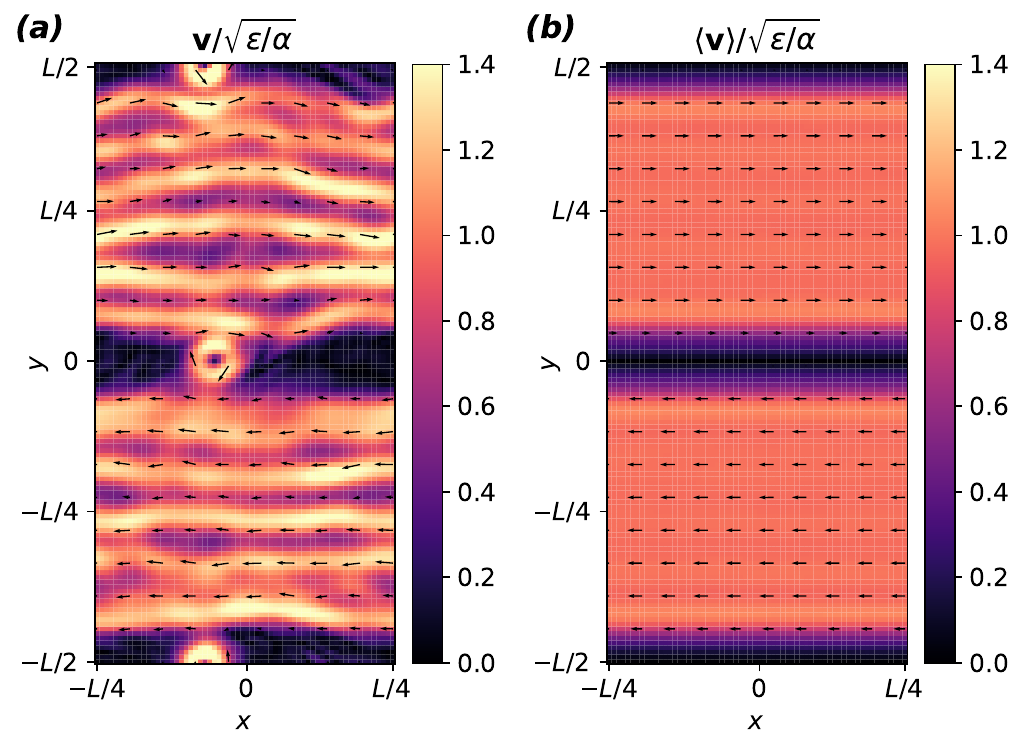}
\caption{\label{fig:Snapshot} LQG jet condensate, showing the velocity  $\boldsymbol{v}=\hat{z}\times\boldsymbol{\nabla}\psi$ snapshot (a) and mean over time (b). The color corresponds to the velocity magnitude. (Simulation-B)}
\end{figure}

The resulting condensate, with two alternating jets along the short side ($x$ direction) of the domain, is shown in Fig.~\ref{fig:Snapshot}. Between the jets, there are two small vortices, similarly to what was found in 2DNS~\cite{frishman_jets_2017}, possibly due to instabilities of the mean flow. In the jet region, the flow is statistically homogeneous in $x$. Small magnitude oscillations along the $y$ direction of the jet amplitude can also be seen in Fig.~\ref{fig:Snapshot}. In steady state, no significant drift of the profile is observed over time, so the mean profile is simply determined from the average of the snapshots without shift. After averaging, we set the axis such that the mean velocity is zero on the $y=0$ line with $U>0$ above it and $U<0$ below it.

\begin{table}[h!]
\caption{\label{tab:sim-list} Parameters of the DNS runs. All runs are performed with hyper-viscosity $p=7$ on a $64\times128$ grid. The forcing wave number is $k_f$, the drag coefficient $\alpha$, viscosity $\nu$, potential energy injection rate $\epsilon$, the ratio of large eddy turnover time and dissipation time scales is $\delta = \alpha \epsilon^{-1/3} L^{2/3}$, the ratio of viscous and forcing time scales is $Re = l_f^{2p-8/3}\epsilon^{1/3}/\nu$ and $T_L$ is the simulation time in units of large eddy turnover time $\tau_L = \epsilon^{-1/3}L^{8/3} $.}
\begin{ruledtabular}
\begin{tabular}{lccccccr}
\textrm{}&
\textrm{ $k_f$ }&
\textrm{ $\alpha{\times 10^{3}}$ }&
\textrm{ $\nu{\times10^{19}}$ }&
\textrm{ $\epsilon {\times 10^{4}}$ }&
\textrm{ $\delta$ }&
\textrm{ $Re{\times10^{-13}}$}&
\textrm{ $T_L$ }\\
\colrule
A & 13 & 2.0 &  7.30  &  2.44 & .109 &  2.26 &  872.9 \\
B & 13 & 1.0 &  7.30  &  2.42 & .055 &  2.25 & 1031.7 \\
C & 13 & 0.5 &  7.30  &  2.42 & .027 &  2.25 & 1122.8 \\
D & 13 & 1.0 &  7.30  &  1.04 & .072 &  1.70 &  185.0 \\
E & 13 & 1.0 &  0.50  &  2.59 & .053 & 33.64 &  176.0 \\
F & 10 & 2.0 & 10.00  &  1.35 & .133 & 26.47 &  191.6 \\
G & 10 & 1.0 & 10.00  &  1.37 & .066 & 26.60 &  155.1 \\
H & 10 & 0.5 & 10.00  &  1.36 & .033 & 26.54 &  252.7 \\
I & 15 & 1.0 &  7.30  &  2.48 & .054 &  0.45 &  327.9 \\
J & 15 & 2.0 &  0.50  &  2.71 & .105 &  6.75 &  300.2 \\
K & 15 & 1.0 &  0.50  &  2.74 & .052 &  6.77 &  206.6

\end{tabular}
\end{ruledtabular}
\end{table}

To obtain a statistical description of the steady state LQG jet condensate we decompose the flow into the mean $\Psi = \langle \psi \rangle$ and fluctuations $\psi^\prime = \psi - \Psi$ focusing on the jet region, where the flow is statistically homogeneous in $x$, and the mean flow depends on $y$ only. The mean flow $\partial_y \Psi\equiv-U(y)$ and the mass flux $\langle v^{\omega'}_y\psi'\rangle$ can be obtained from the mass flux balance (average of equation \eqref{eq:LQG}) and the potential energy balance, neglecting kinetic energy dissipation for the fluctuations and cubic-in-fluctuations terms. The latter assumes that non-linear interactions are dominated by mean-flow-turbulence interactions at the relevant scales, and is also known as the quasi-linear approximation, see~\cite{marston_recent_2023} for a review. The derivation and the comparison to DNS are presented in~\cite{SI}, and here we only cite the resulting leading order solution
\begin{eqnarray}
\partial_{y}\Psi =-U= \pm\sqrt{\frac{\epsilon}{\alpha}}, \label{eq:JetSol U}
\\
\label{eq:JetSol pv}
\langle\psi^{\prime}v_{y}^{\omega\prime}\rangle	= \pm\sqrt{\epsilon \alpha}.
\end{eqnarray}
In agreement with~\eqref{eq:JetSol U}, the simulated mean velocity $U(y)$ is indeed constant in the region where the jets are strong, rapidly switching sign in a thin transition region between the jets, as can be seen in Fig.~\ref{fig:Snapshot}(b).

\section{Second order statistics: two-point correlation functions}
\subsection{Analytical results}
Given an expression for the mean flow, we are now in a position to go further in the perturbation theory and consider the full second order (single-time) statistics. It is sufficient to consider
the two-point correlation function $\left\langle \psi_{1}^{\prime}\psi_{2}^{\prime}\right\rangle \equiv\left\langle \psi^{\prime}(\boldsymbol{r}_{1})\psi^{\prime}(\boldsymbol{r}_{2})\right\rangle $
where $\boldsymbol{r}_{i}=(x_{i},y_{i})$, from which other single and two-point second-order correlation functions can be subsequently derived. To obtain an expression for $\left\langle \psi_{1}^{\prime}\psi_{2}^{\prime}\right\rangle $, we will use that in a statistically steady state
\begin{equation}
0=\partial_{\tau}\left\langle \psi_{1}^{\prime}\psi_{2}^{\prime}\right\rangle =\sum_{i\neq j}\left\langle \psi_{j}^{\prime}\partial_{\tau}\psi_{i}^{\prime}\right\rangle. \label{eq:2PT der dt psi-psi =00003D 0}
\end{equation}
The evolution equation for the fluctuations $\partial_{\tau}\psi_{i}^{\prime}=\partial_{\tau}\left(\psi-\Psi\right)$
is obtained by subtracting the average of equation  \eqref{eq:LQG} from \eqref{eq:LQG}, giving
\begin{multline}
    \partial_{\tau}\psi^\prime = -\bm{v}^{\omega}\cdot\bm{\nabla}\psi+\partial_{y}\left\langle v_{y}^{\omega\prime}\psi^{\prime}\right\rangle \\
    +f+\alpha\nabla^{2}\psi^{\prime}-\nu(-\nabla^{2})^{p}\psi^{\prime},
\end{multline}
where we have used that $V_{y}^{\omega}\equiv\partial_{x}\left\langle \omega\right\rangle =0$ (due to homogeneity in $x$).  Evaluating the derivative at point $\boldsymbol{r}_{i}$, multiplying by $\psi_{j}^{\prime}$ and averaging gives
\begin{multline}
    \left\langle \psi_{j}^{\prime}\partial_{\tau}\psi_{i}^{\prime}\right\rangle =-\left\langle \psi_{j}^{\prime} \bm{v}^{\omega}_i\cdot\bm{\nabla}\psi_{i}\right\rangle +\left\langle f_{i}\psi_{j}^{\prime}\right\rangle
    \\ +\alpha\left\langle \psi_{j}^{\prime}\nabla^{2}\psi_{i}^{\prime}\right\rangle -\nu\left\langle \psi_{j}^{\prime}(-\nabla^{2})^{p}\psi_{i}^{\prime}\right\rangle,
\end{multline}
where $\bm{v}^{\omega}_i \equiv \bm{v}^{\omega}(\bm{r}_i)$. Note that no summation over $i$ is implied here. The cubic term reads
\begin{multline}
\left\langle \psi_{j}^{\prime}\bm{v}_{i}^{\omega} \cdot\bm{\nabla}\psi_{i}\right\rangle =\partial_{y}\Psi_{i}\left\langle \psi_{j}^{\prime}v_{y}^{\omega\prime}(\mathbf{r}_i)\right\rangle \\ +V^{\omega}_x(\mathbf{r}_i)\langle \psi'_j\partial_{x}\psi'_i\rangle+\left\langle \psi_{j}^{\prime}\bm{v}_{i}^{\omega\prime}\cdot\bm{\nabla}\psi_{i}^{\prime}\right\rangle,
\end{multline}
again using that $V_{y}^{\omega}=0$. As the derivatives act on $\boldsymbol{r}_{i}\neq \boldsymbol{r}_{j}$ we can take them out of the average, resulting in
\begin{equation}
\begin{split}
&\left\langle \psi_{j}^{\prime}\partial_{\tau}\psi_{i}^{\prime}\right\rangle =\\&-\left\{ \partial_{y}\Psi_{i}\nabla_{i}^{2}\partial_{x_{i}}+V^{\omega}_x(\mathbf{r}_i)\partial_{x_i}-\alpha\nabla_{i}^{2}+\nu(-\nabla_{i}^{2})^{p}\right\} \left\langle \psi_{j}^{\prime}\psi_{i}^{\prime}\right\rangle
\\& +\left\langle f_{i}\psi_{j}^{\prime}\right\rangle -\bm{\nabla}_i\cdot\left\langle \bm{v}^{\omega\prime}_i\psi_{j}^{\prime}\psi_{i}^{\prime}\right\rangle. \label{eq:2PT der psi dt psi}
\end{split}
\end{equation}
where $\bm{\nabla}_i$ and $\nabla_{i}^{2}$  denotes the gradient and Laplacian
with respect to $\boldsymbol{r}_{i}$.
Finally using (\ref{eq:2PT der dt psi-psi =00003D 0}) and (\ref{eq:2PT der psi dt psi})
we get
\begin{equation}
\begin{split}
\sum_{i=1,2}&\left\{ \partial_{y}\Psi_{i}\nabla_{i}^{2}\partial_{x_i}+V^{\omega}_x(\mathbf{r}_i)\partial_{x_i}-\alpha\nabla_{i}^{2}+\nu(-\nabla_{i}^{2})^{p}\right\} \left\langle \psi_{1}^{\prime}\psi_{2}^{\prime}\right\rangle \\=&
2\chi_{12}-\bm{\nabla}_1 \cdot \left\langle \bm{v}_1^{\omega\prime}\psi_{1}^{\prime}\psi_{2}^{\prime}\right\rangle -\bm{\nabla}_2 \cdot \left\langle \bm{v}_2^{\omega\prime}\psi_{1}^{\prime}\psi_{2}^{\prime}\right\rangle,
\end{split}
\label{eq:JET 2-pt function}
\end{equation}
where we have used that the force two-point
correlation function is given by $\left\langle f(\boldsymbol{x}_{1},t)f(\boldsymbol{x}_{2},t)\right\rangle =2\chi_{12}\delta(t-t^{\prime})$.

\begin{figure*}[t h]
\includegraphics[width=2\columnwidth]{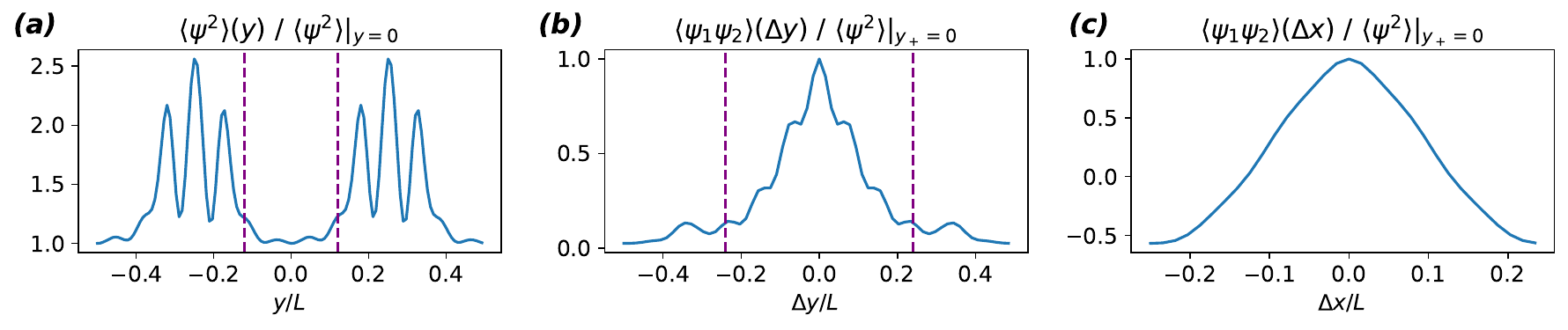}
\caption{\label{fig:2-pt_1D} The two-point correlation function $\langle \psi_1^\prime \psi_2^\prime \rangle$ as measured in DNS (Simulation-B) with (a) $\Delta x = \Delta y = 0$ (b) $\Delta x = y_+ = 0$ and (c) $\Delta y = y_+ = 0$. The region where the leading order solution for the mean flow applies is delimited by dashed lines. (Simulation-B)}
\end{figure*}

Having derived the general equation for the two-point correlation function, equation \eqref{eq:JET 2-pt function}, we will proceed using a perturbative approach. For the mean flow, we will use the leading order solution cited above. In the same manner as was done for the single point correlation function, we shall neglect the viscous dissipation term (there is no dissipative anomaly) as well as that by linear friction, since in the condensate regime we expect the dissipation of the fluctuations of $\psi$ to be a sub-leading effect, e.g.
\begin{equation}
\frac{\alpha\nabla_{i}^{2}\left\langle \psi_{1}^{\prime}\psi_{2}^{\prime}\right\rangle }{\partial_{y}\Psi_{i}\nabla_{i}^{2}\partial_{x_{i}}\left\langle \psi_{1}^{\prime}\psi_{2}^{\prime}\right\rangle }\sim\frac{\alpha^{3/2}}{\sqrt{\epsilon}}l_{2}= \delta^{3/2} \frac{l_2}{L} \leq \delta^{3/2} \ll 1,
\end{equation}
where $l_{2}\le L$ is the length scale of the two-point function and assuming the condensate regime with $\delta \ll 1$.
We will further use the quasi-linear approximation, expecting that at leading
order the cubic fluctuation terms are negligible compared to the mean-flow-fluctuations term  $\partial_{y}\Psi_{i}\nabla_{i}^{2}\partial_{x_{i}}\left\langle \psi_{1}^{\prime}\psi_{2}^{\prime}\right\rangle$.
This gives
\begin{equation}
\left\{ \partial_{y_{1}}\Psi_{1}\nabla_{1}^{2}\partial_{x_{1}}+\partial_{y_{2}}\Psi_{2}\nabla_{2}^{2}\partial_{x_{2}}\right\} \left\langle \psi_{1}^{\prime}\psi_{2}^{\prime}\right\rangle =2\chi_{12}.
\end{equation}
Using homogeneity in the $x$ direction (also for $\chi_{12}$ which
only depends on $x_{1}-x_{2}$), i.e. that $\partial_{x_{1}}=-\partial_{x_{2}}$
when acting on the two-point function, and $\partial_y\Psi=\sqrt{\epsilon/\alpha}$ at leading order, simplifies the advection operator to $\nabla_{1}^{2}\partial_{x_{1}}+\nabla_{2}^{2}\partial_{x_{2}}=(\partial_{x_{1}}^{2}+\partial_{y_{1}}^{2})\partial_{x_{1}}-(\partial_{x_{1}}^{2}+\partial_{y_{2}}^{2})\partial_{x_{1}}=(\partial_{y_{1}}^{2}-\partial_{y_{2}}^{2})\partial_{x_{1}}$.
The equation then reads
\begin{equation}
\left(\partial_{y_{1}}+\partial_{y_{2}}\right)\left(\partial_{y_{1}}-\partial_{y_{2}}\right)\partial_{x_{1}}\left\langle \psi_{1}^{\prime}\psi_{2}^{\prime}\right\rangle =2\sqrt{\frac{\alpha}{\epsilon}}\chi_{12}.
\end{equation}
Changing variables to $y_{+}=(y_{1}+y_{2})/2$ and $y_{-}=(y_{1}-y_{2})/2=\Delta y/2$
so that $\partial_{y_{+}}=\partial_{y_{1}}+\partial_{y_{2}}$ and
$\partial_{y_{-}}=\partial_{y_{1}}-\partial_{y_{2}}$, finally gives the equation for the two-point function in compact form
\begin{equation}
\partial_{y_{+}}\partial_{y_{-}}\partial_{x_{1}}\left\langle \psi_{1}^{\prime}\psi_{2}^{\prime}\right\rangle =2\sqrt{\frac{\alpha}{\epsilon}}\chi_{12}.\label{eq:2-pt eq}
\end{equation}
We now briefly outline the solution of equation~\eqref{eq:2-pt eq}, leaving the detailed derivation to Appendix~\ref{appendix: sol_2pt}.
The solution will be a sum of the particular and the homogeneous solutions of equation~\eqref{eq:2-pt eq}. We begin with the former, first noting that the forcing correlation function $\chi_{12}$ in equation~\eqref{eq:2-pt eq} should be replaced by $\tilde{\chi_{12}}$ 
\begin{equation}
    \tilde{\chi}_{12}=\chi_{12}-\int_{-\frac{L_y}2}^{\frac{L_y}2}
    \frac{ds}{L_y}
    \chi_{12}(\Delta x, s)-\int_{-\frac{L_x}2}^{\frac{L_x}2}
    \frac{ds}{L_x}\chi_{12}(s, \Delta y).
\end{equation} where the $\Delta x$ and $\Delta y$ independent parts (the respective $k_x=0$ and $k_y=0$ Fourier modes) are subtracted. This is necessary as these modes do not satisfy the Fredholm alternative, so the particular solution for them must be determined at next order, see Appendix~\ref{appendix: sol_2pt}.
 The modified equation can now be straightforwardly integrated to obtain the particular solution. Note that for small separations $\Delta x,\Delta y \ll l_f$ $\tilde{\chi}_{12}\approx \chi_{12}$, while for $\Delta x,\Delta y \gg l_f$ $\tilde{\chi}_{12}\ll 1$ so  the influence of the forcing is limited to scales $\Delta x,\Delta y < l_f$, see Appendix~\ref{appendix: sol_2pt}. 
 
While the forcing provides the leading order contribution to the odd in $\Delta x$ part of the correlation function, corresponding to parity+time reversal symmetry breaking, the even contribution at leading order must come from the homogeneous solutions to~\eqref{eq:2-pt eq}. Those are the zero modes of the advection operator $\mathcal{L}_1+\mathcal{L}_2=\nabla_{1}^{2}\partial_{x_{1}}+\nabla_{2}^{2}\partial_{x_{2}}=\partial_{y_{+}}\partial_{y_{-}}\partial_{x_{1}}$ :
\begin{equation}
\left\langle \psi_{1}^{\prime}\psi_{2}^{\prime}\right\rangle _{\text{hom}}=C(\Delta y,\Delta x)+C_1(y_{+},\Delta x)+C_2(y_{+},\Delta y).
\end{equation}
The relevant form of the solution in our case is only $C(\Delta y,\Delta x)$ as we detail in Appendix~\ref{appendix: sol_2pt}.
Thus, the full solution reads
\begin{equation}
\begin{split}
\left\langle \psi_{1}^{\prime}\psi_{2}^{\prime}\right\rangle& =C(\Delta y,\Delta x) \\
 & +2y_{+}\sqrt{\frac{\alpha}{\epsilon}}\int_{0}^{\Delta x}dz\int_{0}^{\Delta y/2}dz'\tilde{\chi}_{12}\left(z,z'\right),
 \label{eq: two point inter}
\end{split}
\end{equation}
Note that for the homogeneous part $C(\Delta x,\Delta y)=C(-\Delta x,\Delta y)=C(-\Delta x,-\Delta y)$, where the first equality is a consequence of the invariance with respect to parity ($x\to -x)$ + time reversal ($t\to -t$) (PT) which we expect the zero modes to have, and the second of the exchange symmetry $\bm{r}_{1}\to \bm{r}_{2}$ of the two-point correlation function. In addition, we get the prediction that for $\Delta x=0$, $\Delta y\neq 0$ the correlation function is independent of $y_+$, as confirmed in DNS Fig.~\ref{fig:2-py_f(y+)}. Our approach lacks information about the boundary conditions to be applied and treats the differential operator perturbatively. It is thus unclear how to determine $C(\Delta x, \Delta y)$, and it may require going to the next order in perturbation theory, which is beyond the scope of the present work.

As a consistency check, we can compute the mass flux $\langle v_{y}^{\omega\prime}\psi^{\prime}\rangle$ directly from our result for the two-point function \eqref{eq: two point inter}. In particular, we directly confirm that, being an odd correlator, it is determined by the inhomogeneous solution to the two-point function equation. We shall compute $\left\langle v_{y}^{\omega \prime} (\mathbf{r}_1) \psi_{2}^{\prime}\right\rangle$ and will subsequently merge the two points.

\begin{align}
&\sqrt{\frac{\epsilon}{\alpha}}\left\langle v_{y}^{\omega \prime} (\mathbf{r}_1) \psi_{2}^{\prime}\right\rangle  = \nonumber\\
& = \nabla_1^2\partial_{x_1}\left[2y_+\int_{0}^{\Delta x}dz\int_{0}^{\Delta y/2}dz'\tilde{\chi}_{12}\left(z,z'\right) \right]+\text{\{odd\}}\nonumber \\
& =  \partial_{y_1}^2\left[2y_+\int_{0}^{\Delta y/2}dz'\tilde{\chi}_{12}\left(\Delta x,z'\right) \right]+\text{\{odd\}}\nonumber \\
& =  \tilde{\chi}_{12}(\Delta x, \Delta y) +\text{\{odd\}}
\label{eq:mass_flux_from_2pt}
\end{align}
where $\text{\{odd\}}$ denotes terms odd in $\Delta y$ and $\Delta x$ which will vanish when we take the single point limit $\Delta x, \Delta y \to 0 $. Note that the zero mode indeed produces only odd contributions in $\Delta x$ (since $C(\Delta x, \Delta y)$ is even under $x_1\to -x_1$ while in ~(\ref{eq:mass_flux_from_2pt}) there is an odd derivative with respect to this variable), which do not contribute.
Taking the limit $\Delta x, \Delta y \to 0 $,  $\tilde{\chi}_{12} \to \epsilon$, up to the contribution to the energy injection rate from modes with $k_x=0$ and $k_y=0$, assumed to be $O(l_f/L)$. We thus get the expected result of $\langle v_{y}^{\omega\prime}\psi^{\prime}\rangle = \sqrt{\alpha \epsilon}$.


\subsection{\textbf{Simulation results for the two-point correlation function}}
\begin{figure}[h!]
\includegraphics[width=1\columnwidth]{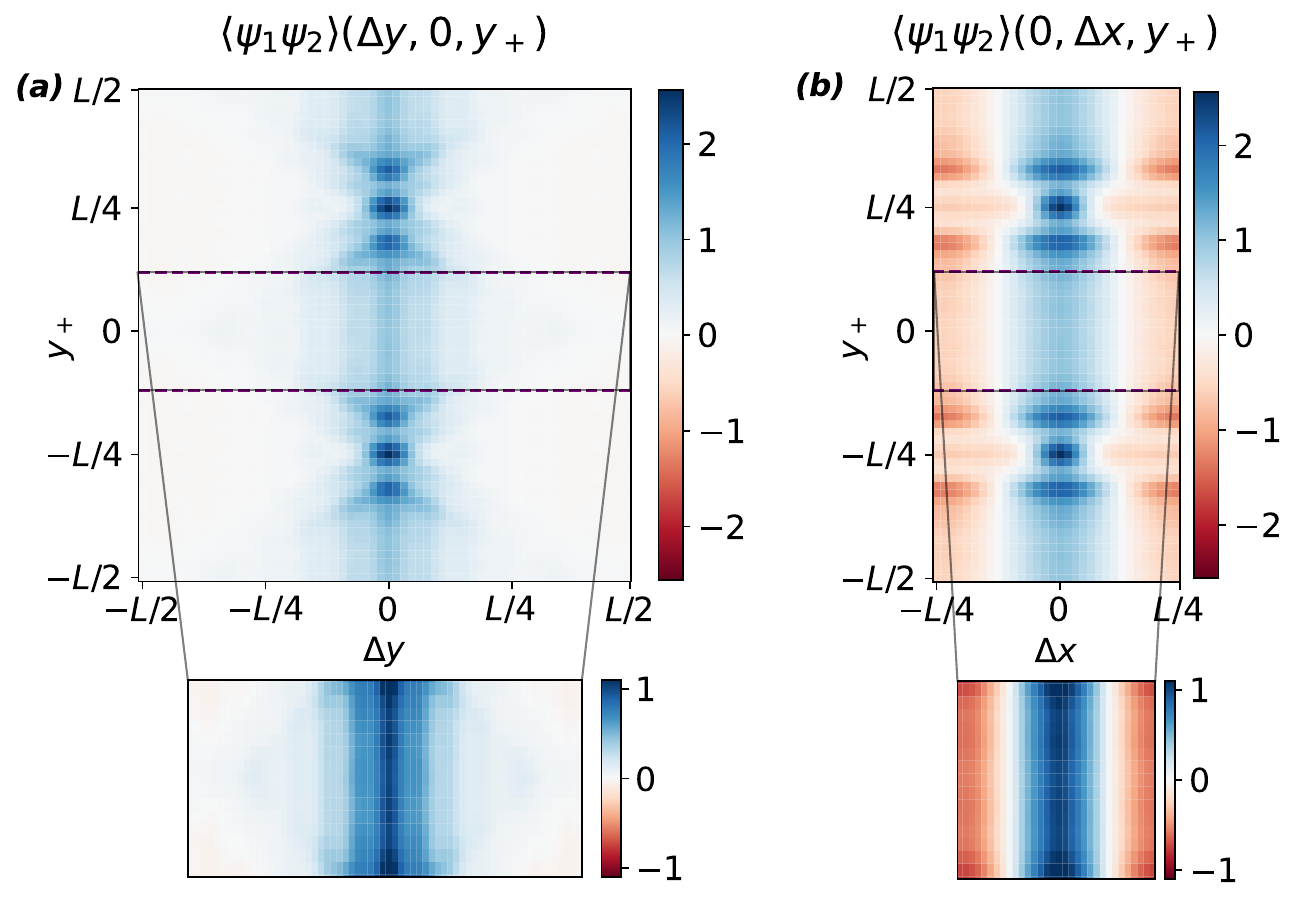}
\caption{\label{fig:2-py_f(y+)} The variation of the two-point function with $y_+$ with: (a) $\Delta x = 0$ and (b) $\Delta y = 0$. (Simulation-B)}
\end{figure}

We now present results from DNS for the two-point function $ \langle \psi_1^\prime \psi_2^\prime \rangle(\Delta y, \Delta x, y_+)$. Note that a-priori the correlation function also depends on $x_+$ (or $x_1$), but taking into account statistical homogeneity in the $x$ direction, we also average over $x_+$ (in addition to time). Using the fact that both jet regions are statistically identical, we compute the two-point function for each of them and average the two to obtain better statistics. In Fig.~\ref{fig:2-pt_1D}(a) we present the variance $\langle \psi'^2\rangle$ as a function of $y$ normalized by its value at $y=0$ at the center of the jet. The jet region, where the leading order solution for the mean flow \eqref{eq:JetSol U} applies, is defined by $|\partial_y U|/\sqrt{\epsilon/\alpha L^2} < 1$ and is delimited by dashed lines, and we expect that $\langle \psi'^2\rangle=C(0,0)$ in this region. The large peaks in the variance $\langle \psi'^2\rangle$ outside the jet region are related to the vortices in between the jets. While they are coherent structures with a large amplitude, which we would normally associate with a mean flow, since they freely move across the domain they contribute to the fluctuations in our averaging procedure. In Fig.~\ref{fig:2-pt_1D}(b) we present the normalized correlation function $\langle \psi_1^\prime \psi_2^\prime \rangle(\Delta y, \Delta x=0, y_+=0)/\langle \psi'^2\rangle(y=0) $, showing how $\psi$ correlations decay with $\Delta y$ when the separation between the points is taken symmetrically around a jets center. In Fig.~\ref{fig:2-pt_1D}(c) we present the normalized correlation function $\langle \psi_1^\prime \psi_2^\prime \rangle(\Delta y=0, \Delta x, y_+=0)/\langle \psi'^2\rangle(y=0) $, showing the correlations in the $x$ direction for points at the center of the jet $y_1=y_2=0$. In Fig.~\ref{fig:2-py_f(y+)} we show that the shape of these correlations is to leading order independent of $y_+$ in the jet region, presenting $\langle \psi_1^\prime \psi_2^\prime \rangle(\Delta y, \Delta x=0, y_+) $ and $\langle \psi_1^\prime \psi_2^\prime \rangle(0, \Delta x, y) $. Note that according to \eqref{eq: two point inter} setting $y_+=0$ or $\Delta x=0$ or $\Delta y=0$, as we do in Fig.~\ref{fig:2-pt_1D}, allows us to probe only the zero mode $C(\Delta x, \Delta y) $ since the inhomogeneous contribution vanishes.

\begin{figure}[h]
\centering
\includegraphics[width=1\columnwidth]{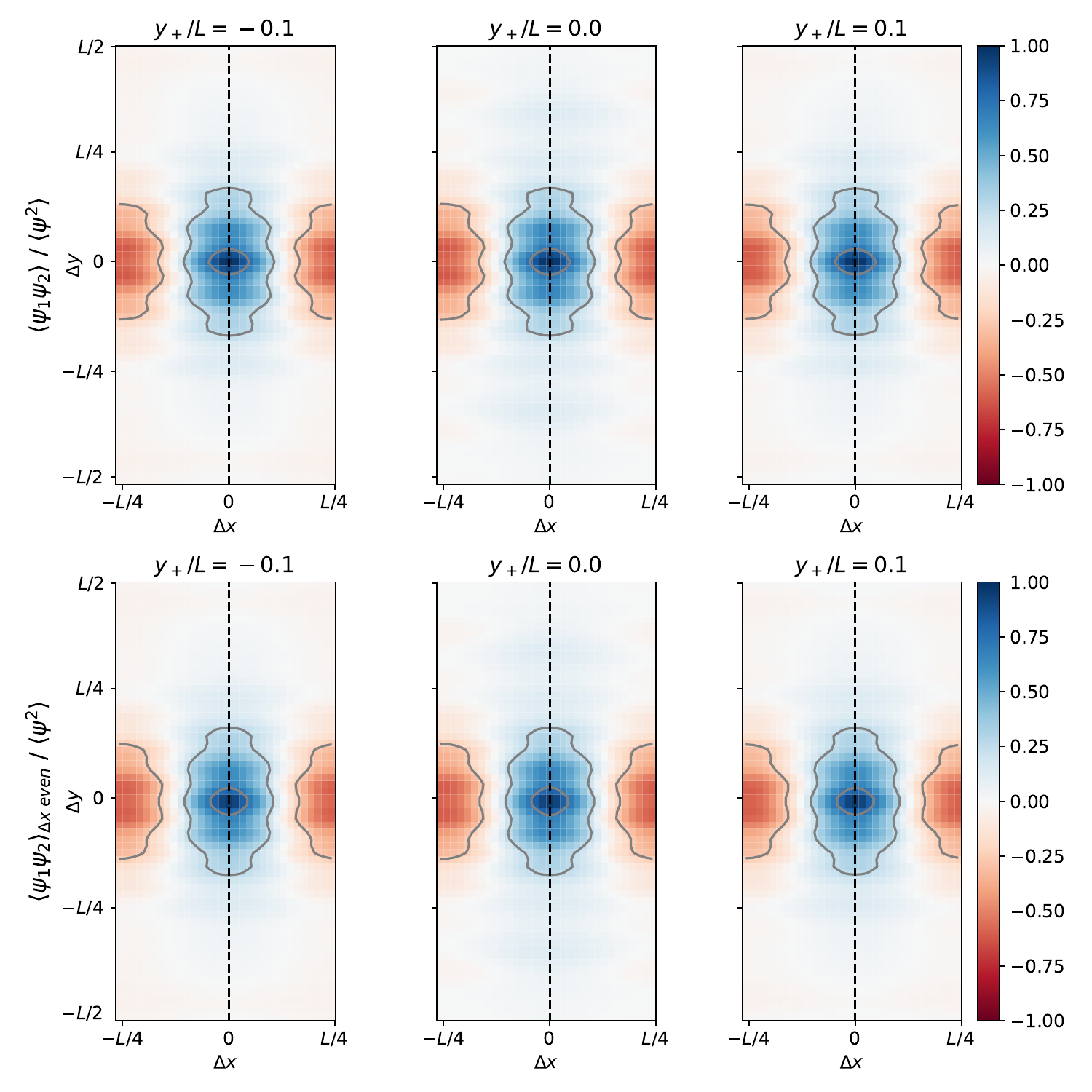}
\caption{\label{fig:C2-2D} The two-point function (averaged over $x_+$) as a function of ($\Delta x$, $\Delta y$) measured at different $y_+$ in the jet region. (Simulation-B)}
\end{figure}

In the upper panel of Fig.~\ref{fig:C2-2D} we present the structure of the correlation function as a function of $\Delta x$ and $\Delta y$ at a few fixed $y_+$. According to \eqref{eq: two point inter} the zero modes $C(\Delta x, \Delta y)$ can be observed by setting $y_+ =0$, as presented in the center panel in Fig.~\ref{fig:C2-2D}. 
Qualitatively, from Fig.~\ref{fig:C2-2D} it appears that the zero mode is symmetric with respect to reflection of $\Delta x$, as expected. To quantify this we decompose the two-point function into its even and odd parts with respect to $\Delta x$ (and separately $\Delta y$) the decomposition given by:
\begin{eqnarray}
    G_{\Delta z \text{ even}}
    &=& \frac{G(\Delta z , ...) + G(-\Delta z , ...)}{2}, \\
    G_{\Delta z \text{ odd}}
    &=& \frac{G(\Delta z , ...) - G(-\Delta z , ...)}{2}.
\end{eqnarray}

To quantify the symmetry of the zero mode we compute the relative power:
\begin{equation}
    R_{\Delta z}\left[G\right] =
    \sqrt{\frac{\iint d^2x G_{\Delta z \text{ odd}}^2}
               {\iint d^2x G_{\Delta z \text{ even}}^2}},
\end{equation}
at $y_+=0$.
For all simulations considered we get that $R_{\Delta x}\left[\langle \psi_1^\prime \psi_2^\prime \rangle \right] \ll 1$ (as well as $R_{\Delta y}\left[\langle \psi_1^\prime \psi_2^\prime \rangle \right]\ll 1$) at $y_+=0$ (and in fact for any $y_+$ inside the jet region). Specifically, in the case of the simulation considered in Figs.~(\ref{fig:2-pt_1D},~\ref{fig:2-py_f(y+)},~\ref{fig:C2-2D}) we get $R_{\Delta x}\left[\langle \psi_1^\prime \psi_2^\prime \rangle \right] = 0.092$ ( $R_{\Delta y}\left[\langle \psi_1^\prime \psi_2^\prime \rangle \right] = 0.099$) at $y_+ =0$. Thus, our results are consistent with the presence of zero modes of the form predicted in \eqref{eq: two point inter}, and support that these zero modes are even with respect to PT as expected from theoretical considerations. 

For $y_+\neq 0$ one expects contributions both from the even and the odd correlators, however, in practice we observe that the two-point function is independent of $y_+$ (as we expect for the zero mode) and looks identical to that at $y_+=0$. Thus it seems that the two-point function is dominated by the zero mode.
In the lower panel of Fig.~\ref{fig:C2-2D} we present the part even with respect to $\Delta x$ of the correlation function. The lower and upper panels indeed appear identical, meaning that the full correlation function is dominated by the even part.
We note that from DNS we also get that $ R_{\Delta y, \Delta x}\left[\langle \psi_1^\prime \psi_2^\prime \rangle \right] = 0.102$, while it should vanish from the exchange symmetry, suggesting that the odd contribution to $\langle \psi'_1\psi'_2\rangle$ is comparable to the numerical noise.


An important question is how does the level of fluctuations scale with the parameters of the problem. In particular, there are at least two small parameters which can be important, $l_f/L$ and $\delta$. In order for the perturbation theory to be consistent, fluctuations should be suppressed compared to the mean flow, the ratio expected to scale as a power of a small parameter. To analyze the scaling of the fluctuations we will focus on single point quantities, but instead of considering $\langle \psi'^2\rangle$ we will consider $\langle u^{\prime 2}\rangle=\langle (\partial_y\psi')^{ 2}\rangle$ and $\langle v'^{ 2}\rangle=\langle (\partial_x\psi')^{ 2}\rangle$. First, since $U=-\partial_y\Psi=\text{Const}$ this will make the comparison to the mean flow much cleaner. In addition, it will allow to compare between the scaling of even under PT correlators and that of the odd correlator $\langle u^\prime v^\prime \rangle = -\langle \partial_x \psi^\prime \partial_y \psi^\prime \rangle$ (while $\langle \psi'^2\rangle$ is an even correlator so there is no odd part to compare to).

For the latter we can derive an analytic expression:
\begin{align}
        \langle v_{y}^{\omega\prime}\psi^{\prime}\rangle &= \langle (\partial_x\nabla^2\psi')\psi'\rangle,
        \nonumber \\
        &= -\langle \partial_x^2\psi' \partial_x \psi'\rangle-\langle \partial_y^2\psi'\partial_x \psi'\rangle,
        \nonumber \\
        &= -\partial_x \frac{\langle (\partial_x\psi')^2\rangle}2 +\langle \partial_y\psi'\partial_x \partial_y\psi'\rangle +\partial_y \langle uv\rangle,
        \nonumber \\
        &=\partial_x\frac{\langle (\partial_y\psi')^2\rangle}2 +\partial_y \langle uv\rangle =\partial_y \langle uv\rangle,
    \label{eq:EP}
\end{align}
where we have used the homogeneity in $x$ repeatedly.
From the leading order solution for the mass flux we thus know that $\partial_y \langle vu\rangle=\sqrt{\alpha\epsilon }$, meaning that $ \langle vu\rangle=\sqrt{\alpha\epsilon } y$ which can be written as $\langle u^\prime v^\prime \rangle  =  (\epsilon L)^{2/3} (y/L) \delta^{1/2}$. This is confirmed by our DNS, shown in Fig.~\ref{fig:ke_flux+uv}(b).

\begin{figure}[h]
\centering
\includegraphics[width=1\columnwidth]{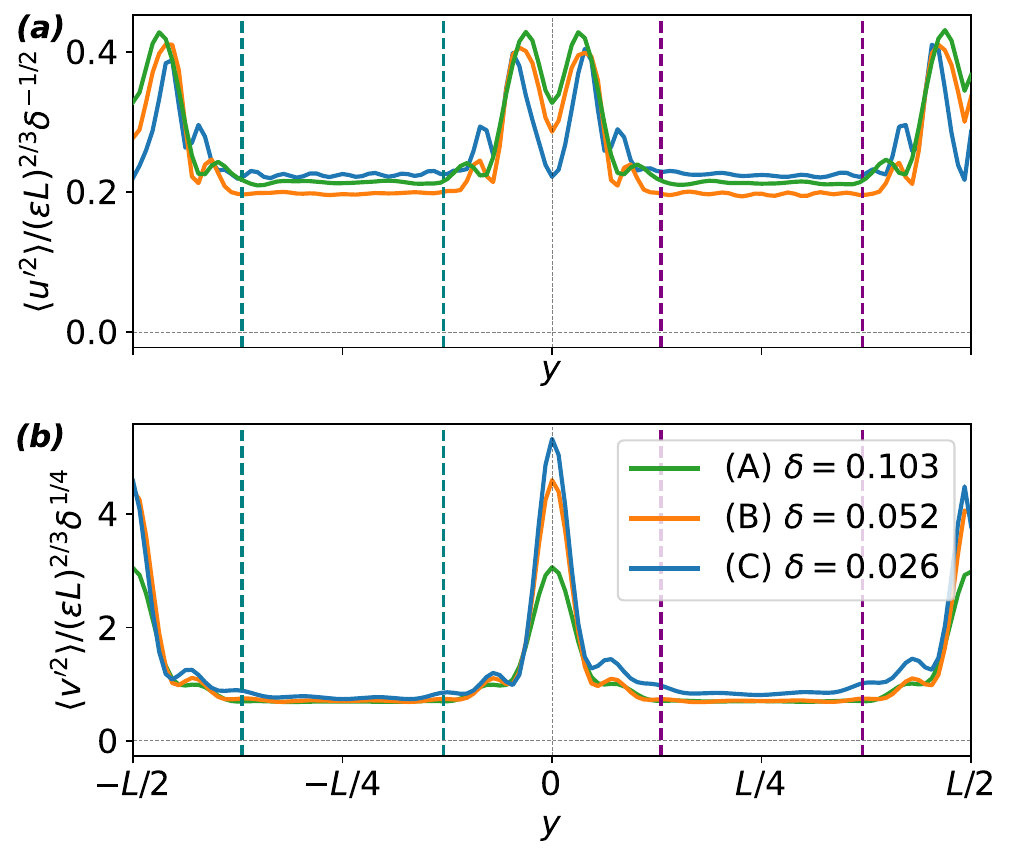}
\caption{\label{fig:u2_and_v2} Rescaled variance of the velocity fluctuations in the direction (a) parallel to the jet ($u^\prime = -\partial_y \psi^\prime$) and (b) perpendicular to the jet ($v^\prime = \partial_y \psi^\prime$), for different values of the parameter $\delta$. The two regions where the leading order solution for the mean flow applies are delimited by vertical dashed lines.}
\end{figure}

On the other hand, we expect the even correlators to be determined by the zero modes, and using equation \eqref{eq: two point inter} we have that $\langle v^{\prime 2}\rangle= -\partial_{\Delta x}^2 C|_{(0,0)} $ and $\langle u^{\prime 2}\rangle= -\partial_{\Delta y}^2 C|_{(0,0)}$ so that both variances are expected to be constant in the jet region. Fig.~\ref{fig:u2_and_v2} confirms this expectation. Note in passing that it is not a-priori clear that we can use \eqref{eq: two point inter} to compute single point correlators of the derivatives of $\psi'$. Indeed, the presence of a direct cascade of the kinetic energy $(\nabla \psi')^2$ implies that non-linear interactions become important for correlators of derivatives of $\psi'$ at small enough distances $\Delta x,\Delta y\ll l_f$, which would invalidate the approximations leading to equation~\eqref{eq:2-pt eq} and its solution \eqref{eq: two point inter} for such correlators. However, as we will see in the next section, in the region where the mean flow is strong the direct cascade is arrested for LQG, which may explain why the solution \eqref{eq: two point inter} can still be used.

\begin{figure}[h]
\centering
\includegraphics[width=1\columnwidth]{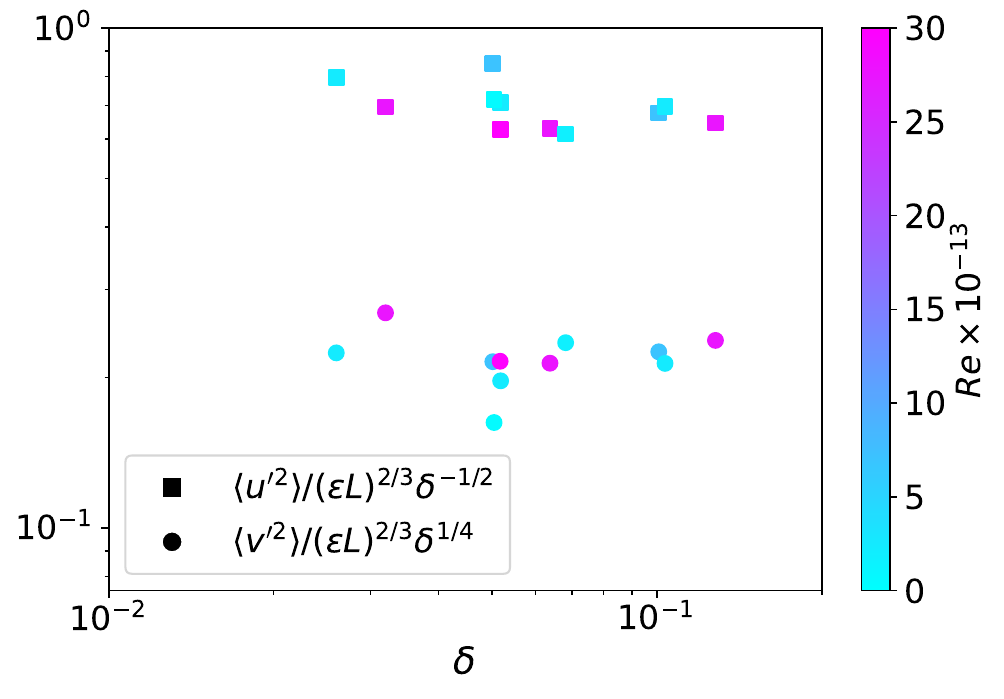}
\caption{\label{fig:scaling} The velocity fluctuation variance inside the jet for the simulations in Table~\ref{tab:sim-list}. Here $u$ is the velocity component parallel to the jet ($x$-component) and $v$ is the velocity component perpendicular to the jet ($y$-component).}
\end{figure}

While lacking a prediction from analytic considerations, we can use the DNS results to determine the scaling of the fluctuations with the parameters of the model. We find that $\langle v^{\prime 2}\rangle$ and $\langle u^{\prime 2}\rangle$ scale differently, probably because of the asymmetry introduced by the mean flow, and we therefore examine them separately.  We find that the following scalings lead to a collapse of data with different run parameters:
\begin{align}
    \langle u^{\prime 2}\rangle
    &\sim(\epsilon L)^{2/3} \delta^{-1/2}, \nonumber\\
    \langle v^{\prime 2}\rangle
    &\sim(\epsilon L)^{2/3} \delta^{1/4}.
    \label{eq:u_v_scaling}
\end{align}
Fig.~\ref{fig:u2_and_v2} demonstrates the collapse of the variance profile when normalized by this scaling for three runs where only $\delta$ is varied.  In Fig.~\ref{fig:scaling} we demonstrate the collapse for runs with varying forcing and viscous scale. Since the velocity variance inside the jet is uniform (Fig.~\ref{fig:u2_and_v2}), here we take the mean value in the middle of the jet as representative for the run.

All in all, we find
\begin{align}
    &\langle u^{\prime 2}\rangle/U^2 \sim \delta^{1/2}, && \langle v^{\prime 2}\rangle/U^2 \sim \delta^{5/4}, \\
    &\langle u^\prime v^\prime \rangle/U^2 \sim (y/L) \delta^{3/2},
\end{align}
using that the mean jet velocity $U^2 \sim (\epsilon L)^{2/3} \delta^{-1}$. Thus, the perturbation theory is indeed consistent, with the fluctuations suppressed compared to the mean flow with powers of $\delta$. It is also worth noting that there isn't one characteristic scaling for the fluctuations, but rather a hierarchy with $\langle u^{\prime 2}\rangle\gg \langle v^{\prime 2}\rangle \gg \langle u^\prime v^\prime \rangle$, in particular the odd correlator is suppressed compared to the even ones. That was also the case for the condensate state in 2DNS~\cite{laurie_universal_2014,frishman_turbulence_2018}. This emphasizes that one cannot straightforwardly use a kinetic theory approach and justify quasi-linear dynamics based on the naive scaling for the fluctuations (coming from the odd correlator).

\section{The direct cascade}
So far we have discussed the potential energy balance and derived the equation for the fluctuations two-point function based on the assumption that potential energy is transferred to large scales, and the only effective way by which it is dissipated is by the formation of the condensate. At the same time, we expect there to be a direct cascade of kinetic energy to small scales and we do not expect the condensate to have a significant influence on this process. Below we will show that it is not the case for LQG turbulence.

\subsection{Spatial kinetic energy balance}
We first derive the spatial kinetic energy balance. To obtain the total kinetic energy balance we act with $(\partial_i \psi) \partial_i$ on \eqref{eq:LQG}. Note that from here, summation is implied over repeated indices. For the non-linear term we have:
\begin{equation}
\begin{split}
& \partial_i \psi \partial_i \partial_j (v_j^{\omega} \psi)
 =
\partial_i  (\partial_i \psi  \partial_j (v_j^{\omega } \psi)) -  \omega \partial_j (v_j^{\omega } \psi)
\\
&= \partial_i \left[\partial_i \psi  \partial_j (v_j^{\omega } \psi)-  \omega  v_i^{\omega } \psi  \right] =\partial_i J_i,
\label{eq:Z_flux}
\end{split}
\end{equation}
 where we have used that since $v_i^{\omega}=\epsilon_{ij} \partial_j \omega$, from symmetry
\begin{equation}
\psi v_i^{\omega} \partial_i \omega
=
 \psi  \epsilon_{ij} \partial_j \omega \partial_i \omega = 0.
\end{equation}
As expected from the inviscid conservation of kinetic energy, this contribution takes the form of a divergence of a flux (of kinetic energy) which we denote by $\bm{J}$. We then decompose the stream-function into its mean and fluctuations $\psi = \Psi + \psi^\prime$, average, and assume homogeneity in $x$ (implying there is only a flux in the $y$ direction):
\begin{eqnarray}
    \langle J_y\rangle &=&   \langle \partial_y \psi  \partial_j (v_j^{\omega } \psi)\rangle - \langle \omega  v_y^{\omega } \psi \rangle  \nonumber\\\
    &=& \partial_y \Psi \partial_y \langle v_y^{\omega \prime} \psi^\prime \rangle +\partial_y \Psi \langle v_y^{\omega\prime} \partial_y \psi^\prime \rangle +\langle v_j^{\omega \prime} \partial_y \psi^\prime  \partial_j \psi^\prime \rangle
    \nonumber \\
     & & -\partial_y^2\Psi\langle v_y^{\omega'}\psi'\rangle-\Psi \langle \omega^\prime  v_y^{\omega \prime}  \rangle -\langle \psi^\prime \omega^\prime  v_y^{\omega \prime}  \rangle.
\end{eqnarray}
The term proportional to $\Psi$ in fact vanishes as
$\langle \omega^\prime v_y^{\omega\prime} \rangle = \langle \omega^\prime \partial_x \omega^\prime \rangle = \partial_x \langle \omega^{2\prime}\rangle/2 =0$.
Thus, the spatial kinetic energy flux is given by:
\begin{equation}
\begin{split}
\langle J_y\rangle =&  \partial_y \Psi \partial_y \langle v_y^{\omega \prime} \psi^\prime \rangle +\partial_y \Psi \langle v_y^{\omega\prime} \partial_y \psi^\prime \rangle
    -\partial_y^2\Psi\langle v_y^{\omega'}\psi'\rangle
    \nonumber \\
      & +\langle v_j^{\omega \prime} \partial_y \psi^\prime  \partial_j \psi^\prime \rangle-\langle \psi^\prime \omega^\prime  v_y^{\omega \prime}  \rangle.
     \end{split}
\end{equation}
It is straightforward to compute the remaining linear terms and the resulting steady state balance of kinetic energy can finally be written as
\begin{equation}
   \partial_{y}\left[\langle J_{y}\rangle+ I_D\right] = \eta - D,
\end{equation}
where $\eta = \langle \partial_i \psi^\prime \partial_i f^\prime \rangle$ is the kinetic energy injection rate, $D$ is the kinetic energy dissipation rate (expected to be mainly due to hyper-viscous dissipation of the fluctuations) and $I_D$ is the flux due to diffusion (e.g. for the drag $I_{D_\alpha}=\alpha \partial_y \langle (\partial_i \psi)^2\rangle/ 2$).

It is also useful to write the kinetic energy balance for the fluctuations. For the non-linear term the contribution can be computed by subtracting $\partial_i\Psi\partial_i\partial_j\langle v_j^{\omega\prime}\psi'\rangle=\partial_y\Psi\partial_y^2\langle v_y^{\omega\prime}\psi'\rangle$ from $\partial_y\langle J_y\rangle$. In particular, for the terms involving the mean flow $\partial_y\Psi$ we have
\begin{equation}
\begin{split}
\partial_y\left[ \partial_y \Psi \partial_y \langle v_y^{\omega \prime} \psi^\prime \rangle+ \partial_y \Psi \langle v_y^{\omega\prime} \partial_y \psi^\prime \rangle \right]-\partial_y\Psi\partial_y^2\langle v_y^{\omega\prime}\psi'\rangle \\
=\partial_y\left[\partial_y \Psi \langle v_y^{\omega\prime} \partial_y \psi^\prime \rangle \right]+\partial_y^2\Psi  \partial_y \langle v_y^{\omega \prime} \psi^\prime \rangle.
\end{split}
\end{equation}
The first term on the bottom line is (part of) a flux of fluctuating kinetic energy, while the second term is (minus) the transfer term of kinetic energy between the fluctuations and the mean flow which we denote by $T$.
We therefore get
\begin{equation}
    \partial_{y}\left[J'_{y} + I_D'\right] = \eta - D' + T,
    \label{eq:k-flac balance}
 \end{equation}
where
\begin{equation}
\begin{split}
    J_y'\equiv \langle J_{y}\rangle -\partial_y \Psi \partial_y\langle v_y^{\omega\prime}  \psi^\prime \rangle
     \end{split}
\end{equation}
is the flux of kinetic energy of the fluctuations, $I_D'$ is the fluctuating flux due to diffusion and $D'$ is the dissipation rate of kinetic energy fluctuations. We expect $T\equiv -\partial_y^2\Psi  \partial_y \langle v_y^{\omega \prime} \psi^\prime \rangle=\partial_y U \partial_y \langle v_y^{\omega \prime} \psi^\prime \rangle\geq0 $ so that the kinetic energy is transferred from the mean flow to the fluctuations. Note that by an order of magnitude estimate we expect the transfer term and the difference between the total flux and the fluctuating flux $\langle J_y\rangle-J'_y$ to be of order $\epsilon/L^2\ll \eta$.

We can now use the leading order solutions \eqref{eq:JetSol U}, \eqref{eq:JetSol pv}, which imply that $\langle \omega \rangle = 0$ and $V_y^\omega = 0$ as well as that $\langle J_y\rangle =J'_y$ and the transfer term vanishes, since both $\partial_y\Psi$ and $\langle v_y^{\omega\prime} \psi^\prime \rangle$ are independent of $y$ to leading order.
Therefore
\begin{equation}
\begin{split}
    \langle J_y\rangle =J_y'&= \partial_y \Psi \langle v_y^{\omega\prime} \partial_y \psi^\prime \rangle
    + \langle v_j^{\omega \prime} \partial_y \psi^\prime  \partial_j \psi^\prime \rangle
     -\langle \psi^\prime \omega^\prime  v_y^{\omega \prime}  \rangle,
     \end{split}
\end{equation}
and the balance for the fluctuations reads
\begin{equation}
\begin{split}
    &\partial_{y}\left[\partial_y \Psi \langle v_y^{\omega\prime} \partial_y \psi^\prime \rangle
    + \langle v_j^{\omega \prime} \partial_y \psi^\prime  \partial_j \psi^\prime \rangle
     +\langle \psi^\prime \omega^\prime  v_y^{\omega \prime}\rangle + I_D'\right]\\
     &= \eta - D'.
     \end{split}
\end{equation}

\begin{figure}[h!]
\includegraphics[width=1\columnwidth]{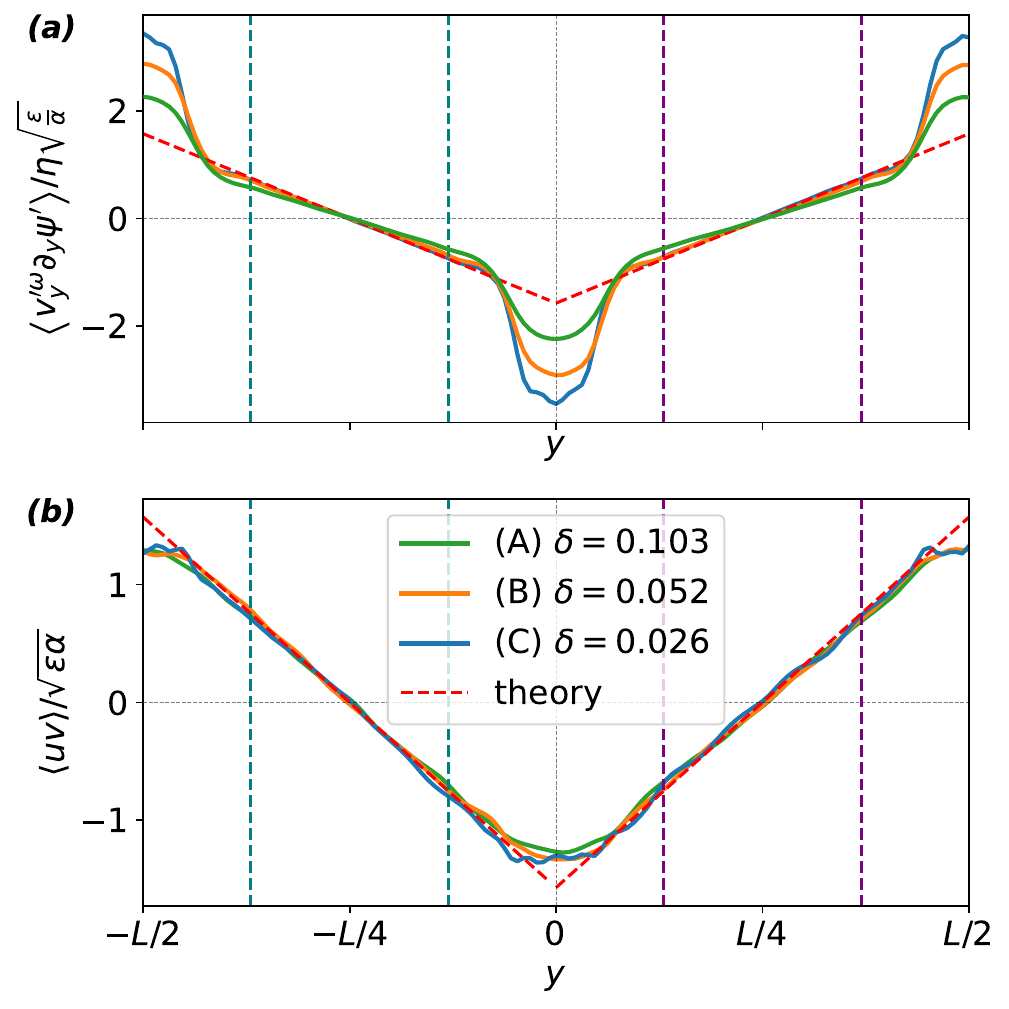}
\caption{\label{fig:ke_flux+uv}
The terms (a) $\langle v_{y}^{\omega\prime} \partial_y \psi^{\prime}\rangle$ and (b) $\langle v u \rangle$ as measured from the DNS (solid lines), rescaled and compared with their theoretical predictions (dashed line), for different values of the parameter $\delta$. The two regions where the leading order solution for the mean flow applies are delimited by vertical dashed lines.}
\end{figure}

We can now directly evaluate the spatial flux of kinetic energy mediated by the mean flow, computing
 $\langle v_{y}^{\omega\prime}\partial_{y}\psi^{\prime}\rangle$ based on our previous results for the two-point function. Indeed, it is given by the limit $\mathbf{r}_2 \to \mathbf{r}_1$ of the two-point function
\begin{equation}
\left\langle v_{y}^{\omega \prime} (\mathbf{r}_1) \partial_{y}\psi_{2}^{\prime}\right\rangle =\left(\partial_{y_{1}}^{2}+\partial_{x_{1}}^{2}\right)\partial_{x_{1}}\partial_{y_{2}}\left\langle \psi_{1}^{\prime}\psi_{2}^{\prime}\right\rangle.
\end{equation}
Note that we only need to consider the inhomogeneous part of the solution \eqref{eq: two point inter} as the zero mode is even under the reflection symmetry $x\to-x$, while there is an odd number of derivative with respect to $x_1$ appearing above. Thus, the contribution from the zero mode will be odd and vanish in the limit $\mathbf{r}_2 \to \mathbf{r}_1$.
Carrying out the calculation, we get
\begin{widetext}
\begin{align}
\label{eq:2p_flux_der_1}
\left\langle v_{y}^{\omega \prime} (\mathbf{r}_1) \partial_{y}\psi_{2}^{\prime}\right\rangle
& = \sqrt{\frac{\alpha}{\epsilon}}\nabla_1^2\partial_{x_1}\partial_{y_2}\left[(y_1 + y_2)\int_{0}^{\Delta x}dz\int_{0}^{\Delta y/2}dz'\tilde{\chi}_{12}\left(z,z'\right) \right]+\text{\{odd\}}\nonumber \\
& = \sqrt{\frac{\alpha}{\epsilon}} \nabla_1^2\partial_{y_2}\left[(y_1 + y_2)\int_{0}^{\Delta y/2}dz'\tilde{\chi}_{12}\left(\Delta x,z'\right) \right]+\text{\{odd\}}\nonumber \\
& =\sqrt{\frac{\alpha}{\epsilon}} \nabla_1^2\left[-\frac{y_1 + y_2}{2} \tilde{\chi}_{12}(\Delta x,\Delta y) + \int_{0}^{\Delta y/2}dz'\tilde{\chi}_{12}\left(\Delta x,z'\right)\right]+\text{\{odd\}}\nonumber \\
& = -y_+\sqrt{\frac{\alpha}{\epsilon}}\nabla_1^2 \tilde{\chi}_{12} +\text{\{odd\}},
\end{align}
\end{widetext}
where $\Delta x=x_1-x_2$, $\Delta y=y_1-y_2$ and we have used that the correlation function is even with respect to $\Delta x\to-\Delta x$, and $\Delta y\to-\Delta y$. In writing \eqref{eq:2p_flux_der_1} we have stated explicitly only the terms which will contribute in the limit $\mathbf{r}_2 \to \mathbf{r}_1$, suppressing the odd contributions. To take the limit we must evaluate $\nabla_1^2 \tilde{\Phi}_{12}$ in this limit, which can be obtained from the definition of the kinetic energy injection rate $\eta$
\begin{equation}
\label{eq:2p_flux_der_2}
\begin{split}
    \eta &= \langle \partial_i f \partial_i \psi\rangle = -\langle f  \nabla^2 \psi \rangle, \\ &=- \lim_{2\to1} \nabla^2_1 \langle \psi_1^\prime f_2\rangle
     = -\lim_{2\to1} \nabla^2_1 \chi_{12}.
\end{split}
\end{equation}
We obtain $\langle v_{y}^{\omega\prime}\partial_{y}\psi^{\prime}\rangle$ by taking $\Delta x, \Delta y \to 0$ of \eqref{eq:2p_flux_der_1} using \eqref{eq:2p_flux_der_2} and get:
\begin{equation}
\langle v_{y}^{\omega\prime}\partial_{y}\psi^{\prime}\rangle =\eta\sqrt{\frac{\alpha}{\epsilon}}y=\frac{\sqrt{\epsilon \alpha}}{l_f^2} y,\label{eq: flux term}
\end{equation}
which is in good agreement with DNS, as presented in Fig.~\ref{fig:ke_flux+uv}(a). Note that here we have assumed that there is a negligible amount of kinetic energy injected into the $k_x=0$ and $k_y=0$ modes (but have not assumed that the forcing is isotropic) so that $\tilde{\eta}=\nabla_1^2 \tilde{\Phi}_{12}\approx\nabla_1^2 \Phi_{12}=\eta$.
As a whole we thus get that at leading order
\begin{equation}
\label{eq:leading_Z_flux}
   \partial_y J_y'=\partial_y \Psi\partial_y\langle v_{y}^{\omega\prime}\partial_{y}\psi^{\prime}\rangle\approx \eta .
\end{equation}
We could have inferred that this flux term will give a contribution of the order of $\eta$ based on an order-of-magnitude estimate. To see this, we first recall that $\langle v_{y}^{\omega\prime}\psi^{\prime}\rangle=\partial_y \langle vu\rangle$, where $u=-\partial_y \psi', v=\partial_x\psi'$. On the other hand, we would like to evaluate $\partial_y\langle v_{y}^{\omega\prime}\partial_y\psi^{\prime}\rangle=-\partial_y\langle v_{y}^{\omega\prime}u\rangle$. As the fluctuations are determined at small scales, we thus expect that derivatives acting on the fluctuating fields inside the average will get a contribution from scales $\sim l_f$, which leads to the estimate $\partial_y\langle v_{y}^{\omega\prime}u\rangle\sim \partial_y \langle vu\rangle/l_f^2\sim \sqrt{\alpha\epsilon }/l_f^2$ in agreement with equation~\eqref{eq: flux term}. Note that the sign of the flux, implying that it carries kinetic energy away from the jet region, seems to be a non-trivial result of the calculation. The direction of the flux is evidently linked to the direction of transfer of potential energy between scales: when potential energy is transferred from the fluctuations to the mean flow, that suppresses the direct cascade in that region, and kinetic energy is carried away from this region.

\begin{figure}[t!]
\includegraphics[width=1\columnwidth]{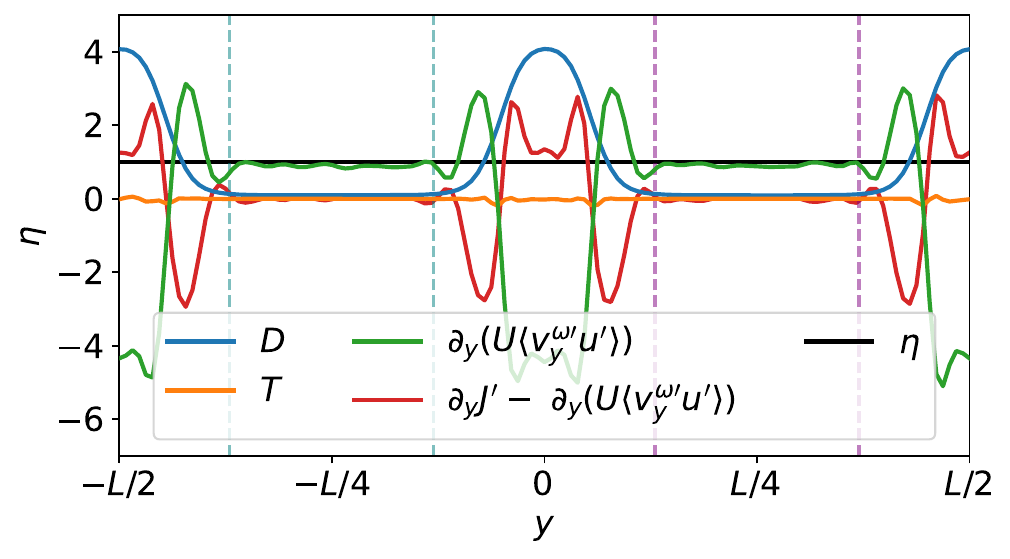}
\caption{Kinetic energy balance of the fluctuations \eqref{eq:k-flac balance}. The dominant flux term in the jet region $U\langle v_y^{\omega\prime} y^\prime \rangle = \partial_y \Psi \langle v_y^{\omega\prime} \partial_y \psi^\prime \rangle$ is plotted separately from the other flux terms. The two regions where the leading order solution for the mean flow applies are delimited by vertical dashed lines.  (Simulation-B)
\label{fig: KE balance}}
\end{figure}

The result for $J_y'$, \eqref{eq:leading_Z_flux}, suggests that all the kinetic energy which is injected locally is carried away by a spatial flux due to the presence of the mean flow. In particular, if there is no spatial flux due to non-linear fluctuations-fluctuations interactions which brings kinetic energy to this region from other regions, this implies that the dissipation of kinetic energy in the jet region is negligible, $D^\prime \ll \eta$.
This is indeed in agreement with our results from DNS, as can be seen from Fig.~\ref{fig: KE balance} where the profiles of the terms in the kinetic energy balance \eqref{eq:k-flac balance} are shown. In the jet region we indeed see that the balance is between the kinetic energy injection and the divergence of the flux $\partial_y (U\langle \langle v_y^{\omega'}u'\rangle)$ due to the mean flow, in accordance with Eq.\eqref{eq:leading_Z_flux}. This implies that the kinetic energy is carried away from the jet region, where the mean flow is strong, before it has time to cascade to small scales and dissipate there --- so that the mean flow effectively arrests the direct cascade. The kinetic energy is then deposited in the region in between the jets where the divergence of this flux becomes negative in most of the region as seen in Fig.~\ref{fig: KE balance}. This is also the region where dissipation of kinetic energy occurs. Note however, that other terms in the flux $J_y'$ also become important in that region, redistributing kinetic energy in the opposite direction to that of $ U\langle \langle v_y^{\omega'}u'\rangle$, as seen in the red curve in Fig.~\ref{fig: KE balance}. In particular cubic terms in fluctuations (not shown separately here) have an important contribution to the flux, which is probably related to the presence of a coherent vortex in that region. 
 

\subsection{Local scale-to-scale flux: the filtering approach}
The presence of a condensate makes our problem inhomogeneous due to the effects of the large-scale mean flow. In the previous section, we have shown indications that this inhomogeneity affects the transfer of kinetic energy to small scales so that we expect the direct cascade to proceed inhomogeneously in space. In this section, we would like to confirm this scenario by directly examining the kinetic energy flux between scales for different regions in the flow. The flux in Fourier space only gives the mean flux for the entire flow, so cannot differentiate between different spatial regions. Instead, we employ a real space filtering technique~\cite{eyink_localness_2009}, combining local spatial information and information about transfer between scales. It relies on a coarse-graining of the fields in real space using a convolution kernel with a characteristic length scale. The convolution with the kernel effectively filters features on scales smaller than its length scale. This approach is somewhat similar to a 2D wavelet transform, that keeps the spatial dependence. It then allows for the decomposition of the fluid kinetic energy (or other quadratic integrals) into band-pass contributions from a series of length scales in real space, writing  the corresponding budget equation gives the transfers of turbulent energy both in space and in scale. The main feature of this approach which will be useful here is the scale-to-scale flux term which is space dependent in this approach. It will allow us to determine the spatial distribution of the flux across scales of potential and kinetic energy. The approach was previously applied to incompressible Navier-Stokes~\cite{eyink_local_1995,chen_physical_2003,chen_physical_2006,eyink_localness_2009}, as well as other kinds of flows, including compressible flows~\cite{aluie_scale_2013}. Here we will perform the scale decomposition and derive the analogous balance equations for the LQG equations.

\begin{figure*}[ht]
\includegraphics[width=2\columnwidth]{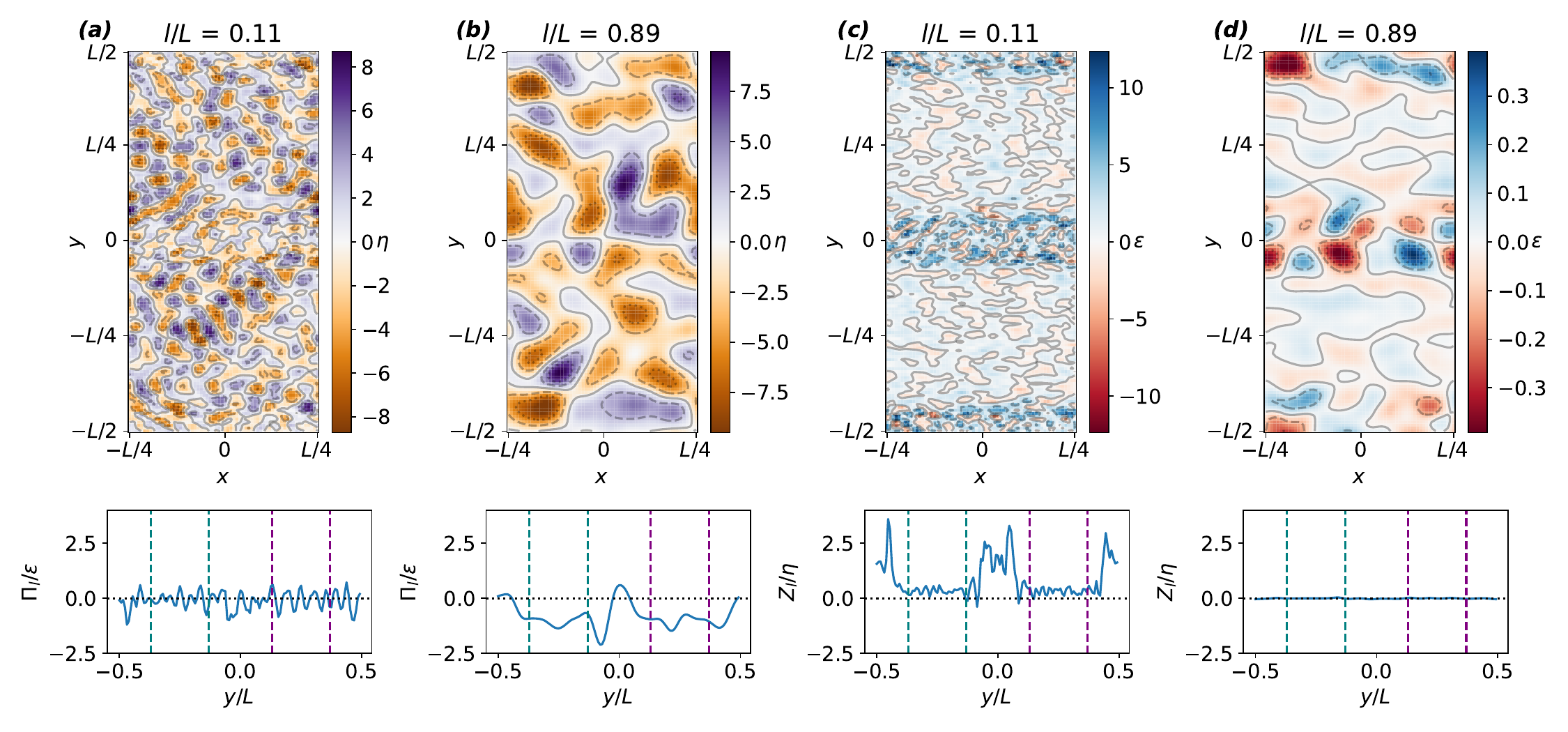}
\caption{\label{fig:filter_snap}
Potential energy flux between scales $\Pi_{l}$ normalized by the total injection rate $\epsilon$ for scales $l/L=0.11$ (a) and $l/L=0.89$ (b). Kinetic energy flux between scales $Z_{l}$ normalized by the total injection rate $\eta$ for scales $l/L=0.11$ (c) and $l/L=0.89$ (d).  Below each plot the average over $x$ is shown. The fields shown are obtained upon taking a short-time average over $15 T_L$ in steady-state. The two regions where the leading order solution for the mean flow applies are delimited by vertical dashed lines. (Simulation-K)}
\end{figure*}

Following \cite{eyink_localness_2009}, we define a smooth low-pass filter as
\begin{equation}
\label{eq:filter}
\overline{\psi}_{l}(\boldsymbol{r})\equiv\int\text{d}\boldsymbol{x}^{\prime}\,G_{l}(\boldsymbol{x}^{\prime})\psi(\boldsymbol{x}^{\prime}+\boldsymbol{r}),
\end{equation}
where the convolution kernel $G_{l}(\boldsymbol{x})$ is taken to be smooth, non-negative, normalized $\int\text{d}\boldsymbol{r}\,G_{l}(r)=1$ and spatially localized. The filter scales with $l$ as $G_{l}(\boldsymbol{r})=l^{-2}G(\boldsymbol{r}/l)$. Specifically, we will choose the Gaussian kernel $G_l(\bm{r}) = e^{-r^2/2 l} / (2\pi l^2)$ when applying filtering to DNS. We may use the filtering operator to write equations for the filtered quantities with a given length scale. In doing so, non-linear interactions cause the emergence of terms representing energy transfer between scales. Filtering is a type of averaging, so the balance equations derived in this section for the large scale kinetic and potential energy are identical in structure to those one derives for the mean flow (where the averaging is over time). In particular, acting with the filter on \eqref{eq:LQG} results in the filtered equations of motion
\begin{equation}
\label{eq:psi_l}
\partial_\tau \overline{\psi_{l}}+\bm{\nabla}\left(\overline{\boldsymbol{v}_{l}^{\omega}}\overline{\psi}_{l}+\bm{\xi}_{l}\right)=\overline{f_{l}}+\alpha\nabla^2 \overline{\psi_{l}}-\nu(-\nabla^2)^{p}\overline{\psi_{l}},
\end{equation}
where $\xi$ is the space dependant flux of the stream function to small scales, defined as
\begin{equation}
\boldsymbol{\xi}_{l}\equiv\overline{\left(\boldsymbol{v}^{\omega}\psi\right)_{l}}-\overline{\boldsymbol{v}_{l}^{\omega}}\;\overline{\psi}_{l}.
\end{equation}
Note that aside from the additional, small scale, spatial flux term $\boldsymbol{\xi}_{l}$, Eq.~\eqref{eq:psi_l} is the same as the regular LQG equation \eqref{eq:LQG} (with $\psi$ replaced by $\bar{\psi}_{l}$).

We are interested in the potential $\overline{e}_{l}\equiv\frac{1}{2}\overline{\psi_{l}}^{2}$  and kinetic $\overline{h}_{l}\equiv\frac{1}{2}\left[\partial_{i}\overline{\psi_{l}}\right]^{2}$ energy balance. We start with the potential energy flux, obtained by multiplying \eqref{eq:psi_l} by $\overline{\psi}_l$ and writing the non-linear terms as a divergence of a flux and a transfer term between scales:
\begin{equation}
    \frac{\partial\overline{e}_{l}}{\partial\tau}+\boldsymbol{\nabla}\cdot\boldsymbol{J}_{l}^{e}=P_{l}^{e}-\Pi_{l}-D_{l}^{e},
\end{equation}
where $\boldsymbol{J}_{l}^{e}$ is a spatial flux term of large-scale energy, $P_{l}^{e}$ is the production of large-scale energy, $D_{l}^{e}$ is the dissipation of energy at large scales and $\Pi_{l}$ is the scale-to-scale energy flux, positive if the transfer is out of the large scales to small scales. The terms are given by:
\begin{eqnarray}
\Pi_{l}&=&-\bm{\nabla}\overline{\psi_{l}}\cdot\bm{\xi}_{l},
\\
\bm{J}_{l}^{e}&=&\overline{\boldsymbol{v}_{l}^{\omega}}\overline{e}_{l}+\overline{\psi_{l}}\boldsymbol{\xi}_{l}-\alpha\bm{\nabla}\overline{e}_{l}
+\nu\bm{I}_{l}^{e,p},
\\
D_{l}^{e}&=&\alpha\left(\partial_i\overline{\psi_{l}}\right)^{2}+\nu\left(\partial_{i_{1}}\cdots\partial_{i_{p}}\overline{\psi_{l}}\right)^{2},
\\
P_{l}^{e}&=&\overline{\psi_{l}}\;\overline{f_{l}},
\end{eqnarray}
where $\boldsymbol{I}_{l}^{e,p}$ is the spatial transport due to hyper-viscosity  $\bm{\nabla}\cdot\bm{I}_{l}^{e,p}\equiv\left[\overline{\psi_{l}}(-\nabla^2)^{p}\overline{\psi_{l}}-(\partial_{i_{1}}\cdots\partial_{i_{p}}\overline{\psi_{l}})^{2}\right]$.

Similarly, to derive the balance for the kinetic energy one takes the derivative of \eqref{eq:psi_l} $\partial_{i}$, and multiplies it by $\partial_{i}\overline{\psi_{l}}$, which gives the equation for low-pass kinetic energy density balance
\begin{equation}
    \frac{\partial}{\partial\tau}\overline{h_{l}}+\boldsymbol{\nabla}\cdot\boldsymbol{J}_{l}^{h}=P_{l}^{h}-Z_{l}-D_{l}^{h},
\end{equation}
where $\boldsymbol{J}_{l}^{h}$ is the spatial flux of large-scale kinetic energy (compare the nonlinear contribution to \eqref{eq:Z_flux}), $P_{l}^{h}$ is the production of kinetic energy at large scales, $D_{l}^{h}$ is the dissipation of kinetic energy at large scales and $Z_{l}$ is the scale-to-scale kinetic energy flux. The different terms are given by:
\begin{eqnarray}
    Z_{l}&=&-(\partial_j  \partial_i \overline{\psi}_{l}) \partial_{i}\xi_{j}, \\
    \boldsymbol{J}_{l}^{h}&=&(\bm{\nabla}\overline{\psi}_{l}) \overline{v_{j}^{\omega}}\partial_j \overline{\psi}_{l}-\overline{\bm{v}_{j}^{\omega}}\overline{\psi}_{l}\nabla^2\overline{\psi}_{l}+\partial_{i}\overline{\psi}_{l}\partial_{i}\bm{\xi}_{l}
    \nonumber \\
    &&-\alpha\boldsymbol{\nabla}\overline{h}_{l}+\nu\bm{I}_{l}^{h,p},\\   D_{l}^{h}&=&\alpha\left(\partial_{i_1}\partial_{i_2} \overline{\psi_{l}}\right)^{2}+\nu\left(\partial_{i_{1}}\cdots\partial_{i_{p+1}}\overline{\psi_{l}}\right)^{2}, \\
    P_{l}^{h}&=&\partial_{i}\overline{\psi_{l}}\partial_{i}\overline{f_{l}},
\end{eqnarray}
with $\bm{\nabla}\bm{I}_{l}^{h,p}\equiv \left[\left(\partial_{i}\overline{\psi_{l}}\right)(-\nabla^2)^{p}\left(\partial_{i}\overline{\psi_{l}}\right)-(\partial_{i_{1}}\cdots\partial_{i_{p+1}}\overline{\psi_{l}})^{2}\right]$.
The transfer terms $\Pi_l$ and $Z_l$ can either be positive, transfering energy from scale $l$ to smaller scales (acting as a sink) or negative, transferring energy from small scales to $l$ (acting as a source). The direct cascade of kinetic energy corresponds to a positive flux (large to small scales), i.e. one expects $Z_l>0$ on average, while an inverse transfer of potential energy implies $\Pi_l<0$ on average. In addition, we expect that $Z_l\approx0$ for large enough scales $l$, as kinetic energy is transferred from the forcing scale to smaller scales, and similarly that $\Pi_l\approx0$ for small enough $l$. The indications above that the direct cascade does not occur in regions where the jets are strong leads to the expectation that $Z_l\approx 0$ in those regions and that the transfer of kinetic energy to small scales is concentrated in between the jets.

We show a short time average over $15 T_L$ of $\Pi_l$ and $Z_l$ in Fig.~\ref{fig:filter_snap}, which demonstrates their spatial distribution. In previous studies of the inter-scale flux in turbulent flows, the flux was observed to be statistically isotropic (as expected) and had regions of both positive and negative contributions on the level of a single snapshot~\cite{rivera_direct_2014}. A definite sign thus emerged only upon averaging. This is roughly what we observe for the potential energy inter-scale flux $\Pi_l$ averaged over short times, Fig.~\ref{fig:filter_snap}(a)(b). In the panel below, we also show the flux when averaged in the $x$ (homogeneous) direction. Then, though the result is still highly fluctuating, a negative flux, on average independent of $y$ in the jet region, emerges for the larger coarse-graining scale $l$. At the small scales, as expected the flux is homogeneous and fluctuating around zero. Note that at the large scales there is an imprint of the jet structure on the flux, with stronger fluctuations in the inter-jet region. 

For the kinetic energy flux $Z_{l}$ the distinction between jet regions and inter-jet regions is evident already at the level of a short-time average, Fig.~\ref{fig:filter_snap}(c)(d). In particular, the flux is visibly suppressed in the jet regions, and for small $l$ is mostly positive between them (rather than having spatially distributed patches of positive and negative contributions of almost comparable magnitude). At large scales $l$ larger fluctuations can be seen in between the jets, but a definite sign is harder to distinguish. These observations are further quantified in the panel below, where upon averaging in the $x$ direction the difference between the two regions is even more clearly seen for the smaller scale $l$. For the larger $l$, the flux fluctuates around zero in the jet region while in between jets a very small negative flux emerges (related to the kinetic energy of the mean flow in that region, which has large gradients there). Thus, we see that most of the direct cascade is indeed concentrated in regions between jets, where the bias between a positive and a negative transfer is so much amplified.

\begin{figure}[h!]
\includegraphics[width=1\columnwidth]{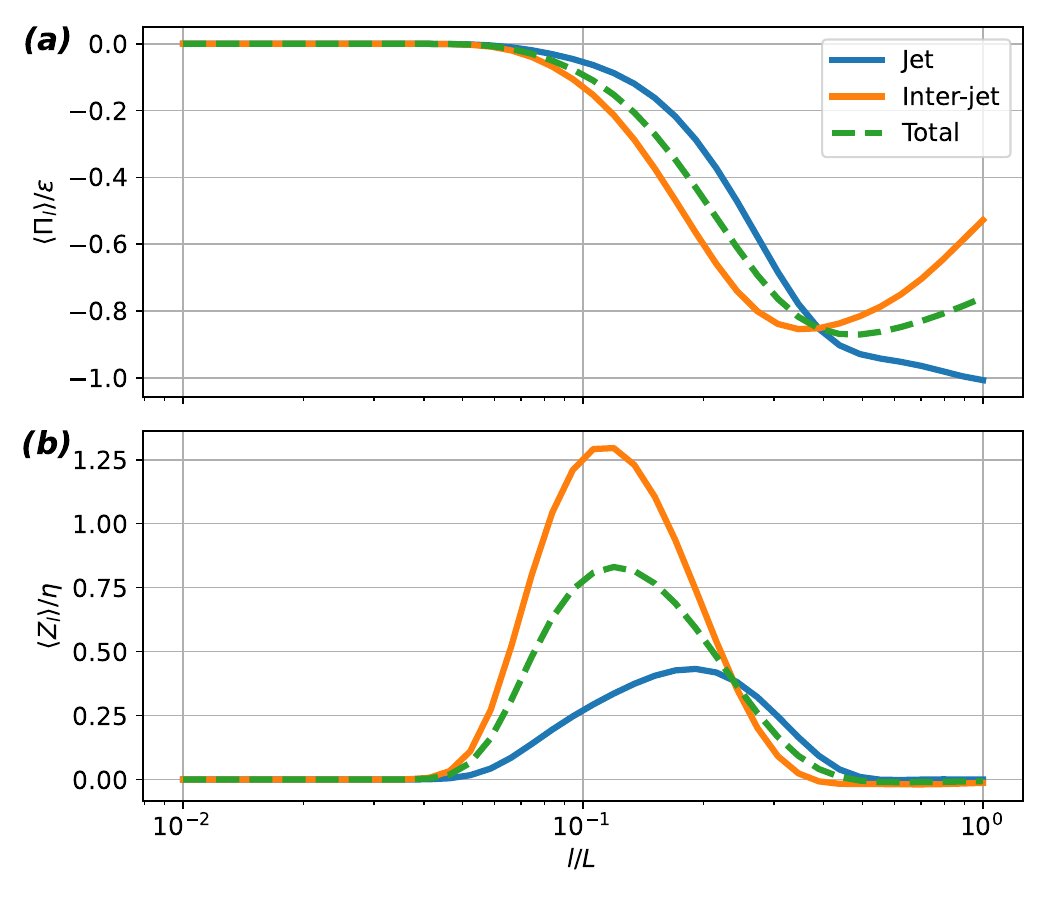}
\caption{\label{fig:filter_flux_decomposition}
Average energy fluxes across length scale $l$ inside (blue) and outside (orange) the jet region. The total flux is denoted by a dashed line (green). (a) Potential energy flux $\left\langle \Pi_{l}\right\rangle$. (b) Kinetic energy flux $\left\langle Z_{l}\right\rangle$.(Simulation-K)}
\end{figure}

Finally, to more systematically quantify the effects observed in Fig.~\ref{fig:filter_snap} we consider the spatially averaged inter-scale fluxes with varying coarse-graining scale $l$. We average both in space and in time, starting once the simulations reach the steady state and up to $100 T_L$. To examine the difference in $Z_{l}$ between spatial regions inside and outside the jets, we split the spatial average of $Z_l(x,y)$ into the jet region $A_{\text{jet}}$ and the region outside the jet $A_{\text{inter-jet}}$. We choose $\text{\ensuremath{A_{\text{jet}}}}$ as the region where the leading order solution for the mean flow applies, as also used in previous figures, and $A_{\text{non-jet}}=A-\text{\ensuremath{A_{\text{jet}}}}$ with $A$ being the entire domain.
We can also define the total average flux (averaged over the whole domain) given by: $ \left\langle Z_{l}\right\rangle =
 (A_{\text{jet}}/A)\left\langle Z_{l}\right\rangle |_{A_{\text{jet}}} +
 (A_{\text{inter-jet}}/A)\left\langle Z_{l}\right\rangle |_{A_{\text{inter-jet}}}$.
For truly homogeneous turbulent flow, the partition would not affect the measurement (as long as both parts are large enough so that the statistics are comparable or the averaging time is long enough). The results are presented in Fig.~\ref{fig:filter_flux_decomposition}. As per our expectation, the potential energy inter-scale flux $\langle \Pi_l\rangle$ is negative everywhere, corresponding to an inverse transfer, while $\langle Z_l\rangle$ is everywhere positive (up to a very slight negative flux for $l/L>0.3$ in the inter-jet region), as expected for a direct cascade. Moreover, the inter-scale flux from small scales $l\lesssim 0.1$ is significantly suppressed in the jet region, implying that so is the direct cascade. This means that the spatial flux $\boldsymbol{J}_l^h$ dominates over the inter-scale flux $Z_{l}$ in the jet region at scales smaller than the forcing scale. Furthermore, at the smallest scales we observe that the total inter-scale flux is completely dominated by the inter-jet region (which occupies a smaller area fraction), in agreement with our observation that the overwhelming majority of the dissipation occurs there, Fig.~\ref{fig: KE balance}. The presence of the mean flow also affects the potential energy inter-scale flux $\langle \Pi_l\rangle$ at large scales $l/L>0.3$, though less dramatically. We observe that the inter-scale flux is reduced in between jets at large enough scales. This is probably a consequence of the inverse transfer being mostly mediated by the mean flow, which takes the form of a vortex in that region. The size of the vortex being of the order of $0.1L$ may thus explain the observed decrease.  

\subsection{Influence of the inverse cascade on the direct cascade in other 2D flows}

We have seen that the condensate has a dramatic effect on the direct cascade in LQG, an effect that appears to be absent in 2DNS~\cite{frishman_jets_2017}. That raises the question of what determines for which types of 2D flows (with an inverse cascade) the latter effect could occur. In particular, recall that LQG and 2DNS are part of a wider class of active scalar equations where a scalar $q$ is advected by a velocity with stream-function $\phi$, with the relation between the two given by $q_{\bm{k}} = |{\bm{k}}|^m \phi_{\bm{k}}$ \cite{pierrehumbert_spectra_1994}. Here $m$ controls the range of the dynamics, for 2DNS, $(m,q,\phi)=(2,\omega,\psi)$ and for LQG $(m,q,\phi)=(-2,\psi,\omega)$ so the velocity is given by derivatives of the scalar, making the dynamics local. All these flows have two positive definite conserved quantities, we shall call $E=\frac{1}2\int q \phi d^2x$ and $Z=\frac{1}2\int q^2d^2x$, where $Z$ cascades to small scales while $E$ is transferred to large scales.

We have seen that for LQG the arrest of the direct cascade occurs in the regions where the mean flow is strong due to a spatial flux of $Z$ away from those regions. A natural question is whether this mechanism could occur for other active scalar flows. To answer this question we consider the balance of $Z$ for the turbulent fluctuations. First note that there are two types of terms involving the mean flow which enter this balance and can interfere with a homogeneous direct cascade: a transfer term between the mean flow and the fluctuations, and a spatial flux term. We expect $Z$ to be transferred to small scales, and therefore an exchange term would tend to remove $Z$ from the mean flow and transfer it to the fluctuations---enhancing the $Z$ injection into the fluctuations in the regions of a strong mean flow. Thus, it is only a spatial flux term which could arrest the direct cascade as in LQG.

We now show that a spatial flux of $Z$ fluctuations due to the mean flow is absent in models where interactions are non-local. In particular, we demonstrate this both for SWQG with $L_d>l_f$ (for $L_d\ll l_f$ the deformation radius influences the direct cascade and we expect a transition to the LQG regime for scales $L_d<l< l_f$), and for an active scalar with $m>0$, assuming the flow is statistically homogeneous in the direction of the mean flow (i.e. that there is no trivial spatial flux due to the inhomogeneity of the turbulence). This is a consequence of the $Z=q^2$ balance for the fluctuations in the steady state:
\begin{equation}
    \partial_i \left\langle u'_i \frac{q'^2}2\right\rangle +\partial_iQ \langle u_iq'\rangle +U_i\partial_i\left\langle \frac{q'^2}2\right\rangle=\eta -D,
\end{equation}
where the third term (which is a spatial flux of $Z'=q'^2$, due to advection by the mean flow) vanishes for a flow statistically homogeneous in the direction of $U$. Thus, the feedback between the condensate and the direct cascade as we have demonstrated in LQG does not exist for an active scalar with $m>0$ which has long-range interactions, but might ,exist in models with $m<0$ where small scale interactions are amplified.

Finally, let us also discuss if the transfer of $Z$ from the mean flow to the fluctuations, $\partial_i Q \langle u_i q'\rangle$ , could significantly enhance the direct cascade in regions of strong mean flow (or large $Q$ gradients). That requires for this term to be of order $\eta$ (the $Z$ injection rate), which we now show is not the case in 2DNS and SWQG. To estimate it let us assume a jet geometry for simplicity, giving  $\partial_yQ \langle u_y q'\rangle \equiv \partial_y Q\langle v q'\rangle $ where we denote $u=-\partial_y\phi', v=\partial_x \phi'$ as we had above. For SWQG (and 2DNS) we have $\langle v q'\rangle=-\partial_y \langle uv\rangle$:
\begin{equation}
    \langle v q'\rangle=\langle \partial_x \psi' (\nabla^2-L_d^{-2})\psi'\rangle =-\langle  \psi' \partial_x\nabla^2\psi'\rangle=-\partial_y\langle uv\rangle,
\end{equation}
where we have already demonstrated the last equality (the Taylor identity) in equation~\eqref{eq:EP} above.
Thus, an order of magnitude estimate provided that $L_d>l_f$ gives $\partial_yQ \langle v q'\rangle\sim U''' \partial_y\langle uv\rangle\sim \epsilon/L^2\ll \epsilon/l_f^2=\eta$ where $\epsilon$ is the injection rate of $E$, meaning that the transfer term is small. This is consistent with the observations in 2DNS~\cite{frishman_jets_2017}, where the cubic-in-fluctuations spatial flux of $Z$ was more significant (though still small) compared to the transfer term.

\section{Discussion}
In this work we have characterized the second order statistics of a jet condensate forming in the large-scale-quasi-geostrophic equation, where potential energy experiences an inverse transfer while kinetic energy cascades to small scales. We have demonstrated that in the regions where the jets are strong the quasi-linear approximation is sufficient to obtain the second-order, two-point correlation functions of all the fluctuating fields ($\psi$ and its derivatives). This is the case since the direct cascade is effectively arrested in those regions, so that non-linear fluctuation-fluctuation interactions are unimportant even for the kinetic energy (and thus can be neglected when determining e.g. correlators of $\nabla \psi$). Using a local coarse-graining approach we have shown that the direct cascade is indeed mostly limited to the inter-jet regions. In the regions where the jets are strong, there is instead a spatial flux of kinetic energy, mediated by the mean flow, which prevents the direct cascade from developing, and which carries the kinetic energy to the inter-jet regions. At the same time, in between the jets we find regions where the quasi-linear approximation for the potential energy necessarily cannot work, since the mean-flow-fluctuations interactions (proportional to $U=-\partial_y\Psi)$ in those regions are small, and there is no other quasi-linear terms which can facilitate a transfer between mean flow and fluctuations. This is a consequence of interactions being local in LQG, so that there are no non-local (e.g. pressure) terms related to the mean flow which can redistribute the energy. Thus we find that in LQG the domain can be decomposed into two distinct regions: one where the dynamics is quasi-linear both for potential energy and for kinetic energy and another where fluctuation-fluctuation interactions overwhelm mean-flow-turbulence interactions for both, which is also where the direct cascade is concentrated. We argue that both phenomena are related to the locality of interactions in LQG, and does not occur for flows with long-range interactions, i.e. an active scalars with $m>0$, as well as models with short-range interactions reaching beyond the forcing scale, namely SWQG with $L_d>l_f$. It remains to be seen if active scalars with $m<0$ or SWQG with $L_d<l_f$, both having dominant interactions below the forcing scale, can exhibit an arrest of the direct cascade as we have found for the limiting LQG case ($m=-2$, $L_d=0$). More generally, understanding the similarities and differences in the condensate state between these two classes of flows away from the LQG limit is an interesting direction for future work.

For the regions where the quasi-linear approximation applies, we have found that fluctuations are suppressed compared to the mean flow with powers of $\delta$, the parameter which quantifies the strength of the condensate. Furthermore, we find that different correlation functions scale differently with $\delta$ and that correlators which are odd with respect to parity+time reversal symmetry are significantly suppressed compared to even correlators. Such a hierarchy was previously observed in 2DNS~\cite{frishman_turbulence_2018}, and points to the fact that constructing a closed perturbative quasi-linear theory for the condensate may be a subtle issue, as it cannot simply rely on a uniform scaling for the fluctuations. Related to this issue, in this work we have determined that even correlators arise from zero modes of an advection operator. We found that these zero modes are homogeneous in the jet region, depending only on $\Delta x$ and $\Delta y$. How exactly those modes are to be determined, including their scaling with $\delta$, however, remains unclear and is left for future work.



\appendix
\section{LQG from SWQG and consistency of limits \label{appendix: LQG_from_SWQG}}
We first briefly remind the physical origin of the shallow water quasi-geostrophic equation, from which the large-scale quasi-geostrophic (LQG) system is derived. It describes a rotating shallow fluid layer, where the horizontal scale of the fluid motion, $L$, is assumed much larger than the layers' mean depth $H$, and which is under the influence of gravity $g$. Assuming a constant rotation rate $\Omega \hat{\bm{z}}$, and a characteristic velocity $U$, the ratio between inertia and the Coriolis force is given by the Rossby number $\text{Ro}=U/\Omega L$. A perturbative expansion in $Ro\ll1$, while assuming $\text{Ro} (L/L_d)^2\sim o(1) $ then gives the SWQG equation~\cite{vallis_atmospheric_2017}
\begin{equation}
    \partial_t q+\bm{v}\cdot \nabla q=\partial_{t}q+J(\psi,q)=0; \quad q=\left(\nabla^{2}-L_{d}^{-2}\right)\psi,
\end{equation}
where $q$ is the potential vorticity, $\psi$ is the stream fucntion with $\bm{v} = \bm{\hat{z} \times \bm{\nabla}}\psi$, $\omega=\nabla^{2}\psi=\left(\boldsymbol{\nabla}\times\boldsymbol{v}\right)\boldsymbol{\hat{z}}$ is the vorticity, $J(\psi,q)$ is the Jacobian operator defined as $J(\psi,q)=\partial_{x}\psi\partial_{y}q-\partial_{y}\psi\partial_{x}q=\epsilon_{ij} \partial_i \psi \partial_j q$ with $\epsilon_{ij}$ the 2D Levi-Civita symbol. The length scale $L_{d}=\sqrt{gH}/2\Omega$ is the Rossby deformation radius. Here, hydrostatic balance relates the variation in the layers' depth $\delta h$ to the pressure $g \rho \bm{\nabla} \delta h = \bm{\nabla}p$. While geostrophic balance relates the stream-function to the pressure so that in total $\psi = (g/\Omega)\delta h$, and there is a single equation for $\psi$.
There are two quadratic invariants in the SWQG system: energy (potential + kinetic) $\int\text{d}^{2}x\,q\psi$ and square potential vorticity $\int\text{d}^{2}x\,q^{2}$. As a consequence, it permits both a direct and an inverse cascade.

Including forcing $f$, friction $\alpha$ (linear drag on velocity) and (hyper) viscosity $\nu$, and using  the box scale $L$ to non-dimensionalize lengths we can write the SWQG equation as
\begin{multline}
     \partial_t \left(\nabla^2-\left(\frac{L}{L_d}\right)^2\frac{1}{L^2}\right)\psi+\epsilon_{ij}\partial_i \psi \partial_j\nabla^2\psi\\
  =f-\alpha \nabla^2\psi+\nu (-\nabla^{2})^p\nabla^2\psi.
\end{multline}
Taking the limit $\left(L_d/L\right)^2\to 0$ formally gives $\partial_t \psi=0$, which is a purely decaying system. Following~\cite{larichev_weakly_1991}, to capture the emerging slow dynamics we will work in rescaled time $\tau = t(L_d/L)^2$ with the limit $L_d/L\to 0$, giving
\begin{multline}
   -\partial_{\tau} \psi+L^2\epsilon_{ij}\partial_i \psi \partial_j\nabla^2\psi\\=f L^2-\alpha L^2 \nabla^2\psi+L^2\nu(-\nabla^{2})^{p+1} \psi.
\end{multline}
Next, we define a new stream-function variable $\tilde{\psi}=L^2\psi$ and a corresponding forcing $\tilde{f}=-fL^2$, drag  $\tilde{\alpha} = \alpha L^2$ and viscosity $\tilde{\nu} = \nu L^2$. With the chosen scaling the relation between the stream-function and the height perturbation in the shallow water system becomes $\tilde{\psi}=(g L^2/\Omega)\delta h$. We thus arrive at the LQG equation~\ref{eq:LQG}.

Let us also demonstrate that the LQG equation can be consistently derived directly from the rotating shallow water equations in the geostrophic limit $\text{Ro}\to 0$.

For a single-layer fluid, and including the Coriolis term, the inviscid shallow water equations (SW) are
\begin{align}
    \p{t} \bm{u} + (\bm{u} \bm{\nabla})\bm{u} + \bm{f}_c \times \bm{u} &= - g\bm{\nabla} \eta, \label{eq:SW1}\\
    \p{t} h + \bm{\nabla}(\bm{u} h) &= 0, \label{eq:SW2}
\end{align}
where $\bm{u}=(u,v)$ is the horizontal velocity, $h$ is the height of the upper free surface (where the bottom surface is assumed flat), $\bm{f}_c=\Omega \hat{\bm{z}}$ and $g$ is gravity.  We apply the geostrophic scaling \cite{vallis_atmospheric_2017} -- assuming that $\bm{u} = (u,v)\sim U$, $(x,y) \sim L$ and an advective time scale $T \sim L/U$. We decompose the free layer height as $h = \overline{h} + \delta h$ with the mean height $\overline{h} = H = \text{const.}$ and the variation $\delta h \sim \Ro H (L/L_d)^2$. The assumptions so far are the same as those made to obtain SWQG. Here, we also rescale the time by $\tau = t(L_d/L)^2$. We will consider two limits, $\Ro \to 0$ and $(L_d/L)\to 0$, and we must specify the relation between them. For what follows we assume that when both limits are taken $\Ro$ tends to zero faster than $(L_d/L)$. Thus we may assume that $(L_d / L) \sim \Ro^b$ with $0< b < 1/2$. Note that with this scaling the height perturbations are still small compared to the mean height as $\delta h/H \sim \Ro^{1-2b} \ll 1$. With this scaling, we obtain the non-dimensional SW momentum equation:
\begin{equation}
    \Ro^{1+2b} \p{\tau} \bm{u}' +
    \Ro (\bm{u}' \bm{\nabla})\bm{u}' + \bm{f}'_c \times \bm{u}' = -\bm{\nabla} \eta',
    \label{eq:tau_sw1}
\end{equation}
and the non-dimensional SW height variation equation:
\begin{multline}
    \Ro \p{\tau}  \delta h' +
    \Ro^{1-2b}(\bm{u'} \bm{\nabla'}) \delta h' \\ +
    (\bm\nabla \cdot \bm{u'})
    \left(1 + \Ro^{1-2b} \delta h'\right) = 0.
    \label{eq:tau_sw2}
\end{multline}
Having expressed both small parameters as a functions of $\Ro$ we expand the velocity $u',v'$ in $\epsilon_i = \epsilon_i(\Ro)$ such that $1 = \epsilon_0 \gg \epsilon_1 \gg ...$ and similarly we expand the height variation $\delta h'$ in $\mu_i = \mu_i(\Ro)$ such that $1 = \mu_0 \gg \mu_1 \gg ...$.
\begin{equation}
    u' = \sum_{i=0}^{\infty} \epsilon_i u'_i, \quad
    v' =  \sum_{i=0}^{\infty} \epsilon_i v'_i, \quad
    \delta h' =  \sum_{i=0}^{\infty} \mu_i \delta h'_i.
    \label{eq:asympSir}
\end{equation}
We leave the asymptotic series arbitrary for now. Substituting the series \eqref{eq:asympSir} into the re-scaled time momentum equation \eqref{eq:tau_sw1} we get:
\begin{align}
    \bm{f}_c' \times \bm{u}_0'
    +O(\epsilon_1; \mu_1; \Ro)
    = -\bm{\nabla} \delta h_0',
\end{align}
where $\bm{f}'_c \equiv f'_0 \hat{\bm{z}} = 1 \hat{\bm{z}}$. The dominant balance (for any $\epsilon_1, \mu_1\ll1$) is between the pressure and Coriolis force thus
\begin{equation}
    f'_0 u'_0 = - \p{y} \delta h_0'; \quad
    f'_0 v'_0 =  \p{x} \delta h_0'\quad
    \Rightarrow \bm{\nabla}\bm{u}'_0 = 0.
    \label{eq:zeroth_order}
\end{equation}
The re-scaled mass conservation \eqref{eq:tau_sw2} gives at leading order the same result. This allows for the definition of the stream function $\psi_0' \equiv \delta h_0'/f_0'$.

Moving on to the next order in perturbation theory to get the dynamics, we
consider the next order of the momentum equation \eqref{eq:tau_sw1}.
\begin{multline}
    \Ro^{1+2b} \p{\tau} \bm{u}'_0 +
    \Ro (\bm{u}'_0 \bm{\nabla})\bm{u}'_0 + \epsilon_1 \bm{f}'_0 \times \bm{u}'_1 \\= -\mu_1 \bm{\nabla} \delta h'_1
    +O(\Ro, \epsilon_2,\mu_2).
    \label{eq:velocity_ord1}
\end{multline}
Taking its curl gives the vorticity $\omega = \nabla \times \bm{u}$ equation
\begin{equation}
    \Ro (\bm{u}'_0 \bm{\nabla})\omega'_0 =
    - f_0' \epsilon_1  (\bm{\nabla} \cdot \bm{u}'_1 )
    +O(\Ro, \epsilon_2,\delta_2),
    \label{eq:omega}
\end{equation}
where the time derivative term has been neglected as it is of higher-order in $\Ro$ than the advection term. Note that the only non-trivial option, in this case, is for $\epsilon_1 = \Ro$ and in general we may assume that $\epsilon_n = \Ro^n$. To proceed consider the next order of \eqref{eq:tau_sw2}. First, we note that \eqref{eq:zeroth_order} gives $(\bm{u'}_0 \bm{\nabla}) \delta h'_0 = 0$ and that $\Ro^{1-2b}  \ll 1$, thus we get
\begin{multline}
    \Ro \p{\tau}  \delta h'_0+
    \Ro^{1-2b}
    \mu_1(\bm{u'}_0 \bm{\nabla}) \delta h'_1
    \\ = -\Ro (\bm\nabla \cdot \bm{u'}_1)
    +O(\mu_1 \Ro ; \Ro^{2-2b}).
    \label{eq:h_ord1}
\end{multline}
Using the $(\bm{\nabla} \cdot \bm{u}'_1 )$ term to relate \eqref{eq:omega} and \eqref{eq:h_ord1} we obtain
\begin{equation}
    \p{\tau} \left(\frac{\delta h'_0}{f_0'} \right) - (\bm{u}'_0 \bm{\nabla})\omega'_0
    =
    \frac{\mu_1}{\Ro^{2b}}
    \frac{1}{f_0'}(\bm{u'}_0 \bm{\nabla})\delta h_1'.
    \label{eq:pre_LQG}
\end{equation}
We wish to obtain a solution for which the leading order velocity does not vanish, thus $\p{\tau}\delta h_0'$ must be determined at this order and $\mu_1 \leq \Ro^{2b}$. If we assume $\mu_1 = \Ro^{2b}\gg \Ro \gg\Ro^{1+2b}$ we obtain from \eqref{eq:velocity_ord1}$ \bm{\nabla} \delta h_{1}' = 0$ and thus $(\bm{u'}_0 \bm{\nabla})\delta h_1' = 0$. Therefore in any case the RHS term can be neglected and equation \eqref{eq:pre_LQG} reduces to
\begin{equation}
    \p{\tau}(\delta h_0' / f_0') - (\bm{u}'_0 \bm{\nabla})\omega'_0
    = 0,
\end{equation}
and with the definitions $\psi_0' \equiv \delta h_0'/f_0'$ and $J(\omega,\psi) = \p{x} \omega \p{y} \psi - \p{y}\omega \p{x} \psi$ we obtain the dimensionless invicid  LQG equation
\begin{equation}
    \p{\tau'}\psi' + J(\omega',\psi')
    = 0.
\end{equation}
Returning the dimensions using $\psi \sim UL$ we have that $\p{\tau}\psi + L^2 J(\omega,\psi)
= 0$. We may absorb the additional factor $L^2$ by redefining the stream function as $\tilde{\psi} = L^2 \psi$ and thus $\tilde{\omega} = \nabla^2 \tilde{\psi} = L^2 \omega$. Dropping the tilde notation we arrive at the LQG equation for the re-scaled stream-function
\begin{equation}
    \p{\tau}\psi + J(\omega,\psi)= 0.
\end{equation}

\section{Solution for the two-point function \label{appendix: sol_2pt}}
Here we describe solution to equation~(\ref{eq:2-pt eq}) for the leading order of the two-point function. The solution to~(\ref{eq:2-pt eq}) is the sum of the solution to the homogeneous equation and the particular solution to the inhomogeneous equation. We begin by describing the latter, assuming the forcing is homogeneous in $y$, so that $\chi_{12}$ depends only on $y_{-}\equiv\Delta y/2$ and $\Delta x$. In principle, in the variables $y_+,y_-,x_1$ the equation can be straightforwardly integrated to obtain the inhomogeneous solution. However, there is a subtle point that has to do with the consistency of the perturbation theory for some of the fluctuations modes.

Consider the Fourier transform of equation~\eqref{eq:2-pt eq} with respect to $\Delta x$ (equivalently $x_1$), denoting the corresponding wavenumber by $k_x$, and with respect to $\Delta y$, denoting the wavenumber by $k_y$. In $\Delta x$, this is possible since the mean flow solution is homogeneous in $x$ and is applicable throughout the $x$ direction, which is periodic. In $\Delta y$, we assume that correlations decay with $\Delta y$ within the region of applicability of the mean flow solution (note that this is not necessarily the case in $y_+$, and that $y_+$ is not a periodic coordinate since the leading order mean flow has a finite range of applicability in $y$). We see that for $k_x=0$ and $k_y=0$ while the left hand side of the equation turns to zero, the right hand side does not. Thus, modes of the forcing with $k_x=0$ or $k_y=0$ need to be treated separately, and equation~\eqref{eq:2-pt eq} is not the leading order equation. This is easily understood for $k_x=0$: perturbations with $k_x=0$ are not advected by the mean flow, so cubic terms or dissipative terms must be important in balancing the injection of the forcing into such modes. That $k_x=0$ and $k_y=0$ modes cannot be treated in a quasi-linear approximation is a completely general statement for mean-flow-turbulence interactions, e.g. also for 2DNS~\cite{frishman_turbulence_2018}. Therefore, in equation~\eqref{eq:2-pt eq} we should subtract these modes from the forcing correlation function:
\begin{equation}
    \tilde{\chi}_{12}=\chi_{12}-\frac{1}{L_y}\int_{-\frac{L_y}2}^{\frac{L_y}2} \chi_{12}(\Delta x, s)ds-\frac{1}{L_x}\int_{-\frac{L_x}2}^{\frac{L_x}2} \chi_{12}(s, \Delta y)ds.
\end{equation}
Then, the inhomogeneous solution reads
\begin{equation}
\left\langle \psi_{1}^{\prime}\psi_{2}^{\prime}\right\rangle {}_{\text{inh}}=(y_{1}+y_{2})\sqrt{\frac{\alpha}{\epsilon}}\int_{0}^{\Delta x}dz\int_{0}^{\Delta y/2}dz'\tilde{\chi}_{12}\left(z,z'\right).
\label{eq:psi_inh_chi}
\end{equation}
In \eqref{eq:psi_inh_chi} we choose the initial point of the integration to be at coincident points $\Delta y=0$, and $\Delta x=0$ which makes the inhomogeneous part symmetric with respect to the replacement $r_{1}\to r_{2}$ (i.e. even under reflection $\Delta x\to-\Delta x,$ $\Delta y\to-\Delta y$). On the other hand, it is odd with respect to $\Delta x\to-\Delta x$ (and $\Delta y\to-\Delta y$) separately. This is what we expect from the fact that the forcing combined with the mean flow break the parity+time reversal symmetry $x\to-x, t\to-t$ of the system, see the discussion in~\cite{SI}.

Let us consider the influence of the subtraction of the modes with $k_x=0$ and $k_y=0$ from the forcing. Assume a typical forcing length scale $l_{f}$, it will be convenient to denote $\chi_{12}(\Delta x,\Delta y)=\epsilon\Phi(\frac{\Delta x}{l_{f}},\frac{\Delta y}{l_{f}})$, such that $\Phi(0,0)=1$. This gives
\begin{equation}
\left\langle \psi_{1}^{\prime}\psi_{2}^{\prime}\right\rangle {}_{\text{inh}}=2y_{+}l_{f}^{2}\sqrt{\alpha\epsilon}\int_{0}^{\frac{\Delta x}{l_{f}}}dz\int_{0}^{\frac{\Delta y}{2l_{f}}}dz'\tilde{\Phi}\left(z,z'\right),\label{eq:psi_inh}
\end{equation}
where
\begin{align}
    &\tilde{\Phi}(z,z')= \nonumber \\
    &=\Phi(z,z')-\frac{l_f}{L_y}\int_{-\frac{L_y}{2l_f}}^{\frac{L_y}{2l_f}} \Phi(z, s)ds
    -\frac{l_f}{L_x}\int_{-\frac{L_x}{2l_f}}^{\frac{L_x}{2l_f}}  \Phi(s, z')ds
    \nonumber \\
    &=\Phi(z,z') - \hat{\Phi}(z, k_y=0)-\hat{\Phi}(k_x=0,z')
\end{align}
with $\hat{\Phi}$ being the Fourier transform of $\Phi$ with respect to $\Delta x$ or $\Delta y$. Here, $\tilde{\Phi}(z,z')$ has no modes with $k_x=0$ or $k_y=0$: $\hat{\tilde{\Phi}}\left(z,k_y=0\right) =\hat{\tilde{\Phi}}\left(k_x=0,z'\right) =0$. For the forcing we have been using in DNS, a direct calculation gives $\hat{\tilde{\Phi}}\left(z,k_y=0\right) =\frac{1}{\pi}\cos 2\pi z$ and $\hat{\tilde{\Phi}}\left(k_x=0,z'\right) =\frac{1}{\pi}\cos 2\pi z'$.

Now, for $\Delta x, \Delta y \leq l_f$ the replacement of $\Phi$ by $\tilde{\Phi}$ does not change the result at leading order: the difference between~\eqref{eq:psi_inh} and the expression when $\tilde{\Phi}(z,z')$ is replaced by $\Phi(z,z')$ is of order $O(l_f/L)$ (after integration). Similarly, we expect that the contribution to two-point correlation functions from $k_x=0$ and $k_y=0$ modes of the forcing, finding which requires a fully non-linear treatment not carried out here, will be small, of order $O(l_f/L)$, compared with the leading order.

On the other hand, for e.g. $\Delta y \approx L/2$ (similarly for $\Delta x\approx L/4$) we notice that
\begin{equation}
\begin{split}
    \int_{0}^{\frac{L_y}{4l_{f}}}dz'\tilde{\Phi}\left(z,z'\right) &\approx  \int_{0}^{\frac{L_y}{2l_{f}}}dz'\tilde{\Phi}\left(z,z'\right)=\frac{1}2\int_{-\frac{L_y}{2l_{f}}}^{\frac{L_y}{2l_{f}}}dz'\tilde{\Phi}\left(z,z'\right), \\
    &=\frac{L_y}{2l_{f}}\hat{\tilde{\Phi}}\left(z,k_y=0\right),
\end{split}
\end{equation}
where we have used that $\tilde{\Phi}$ is a decaying function of $z'$ assuming that $\int_{\frac{L_y}{4l_f}}^{\frac{L_y}{2l_f}}dz'\tilde{\Phi}\left(z,z'\right)\to 0$ as $L_y/l_f\to\infty$, that we are working in the regime $L/l_f\gg1$, and that $\tilde{\Phi}\left(z,z'\right)$ is even in $z'$: $\tilde{\Phi}(z,z')=\tilde{\Phi}(z,-z')$ (corresponding to the assumed statistical reflection symmetry $y\to - y$ of the forcing). We then have
\begin{equation}
\begin{split}
    \int_{0}^{\frac{\Delta x}{l_{f}}}dz\int_{0}^{\frac{L_y}{2l_{f}}}dz'\tilde{\Phi}\left(z,z'\right)\approx \frac{L_y}{2l_f}\int_{0}^{\frac{\Delta x}{l_{f}}}dz\hat{\tilde{\Phi}}\left(z,k_y=0\right) =0.
\end{split}
\end{equation}
Thus, we see that the forcing influences two-point correlation functions only for $\Delta x,\Delta y \leq l_f$, where we can use $\tilde{\Phi}(z,z')\approx \Phi(z,z')$, while for $\Delta x, \Delta y \approx L$ the contribution from the inhomogeneous solution is negligible.

While the forcing provides the leading order contribution to the odd in $\Delta x$ part of the correlation function, corresponding to parity+time reversal symmetry breaking, the even contribution at leading order must come from the homogeneous solutions to~\eqref{eq:psi_inh}. Those are the zero modes of the advection operator $\mathcal{L}_1+\mathcal{L}_2=\nabla_{1}^{2}\partial_{x_{1}}+\nabla_{2}^{2}\partial_{x_{2}}=\partial_{y_{+}}\partial_{y_{-}}\partial_{x_{1}}$ :
\begin{equation}
\left\langle \psi_{1}^{\prime}\psi_{2}^{\prime}\right\rangle _{\text{hom}}=C(\Delta y,\Delta x)+C_1(y_{+},\Delta x)+C_2(y_{+},\Delta y).
\end{equation}
To determine which zero modes contribute to the correlation function we need to take into account the boundary conditions. First, we assume that the fluctuations decorrelate as $\Delta y\to L$, as confirmed in DNS Fig.~\ref{fig:2-pt_1D}(b), implying that $C_1(y_+,\Delta x)=0$. Indeed, the odd and even parts of the correlation function should decay to zero separately in this limit. Also, since $C(\Delta x,\Delta y)$ is independent of $y_+$ while $C_2$ is independent of $\Delta x$, $C_1(y_+,\Delta x)$ should separately decay to zero as $\Delta y\to L$, implying it must be identically zero.
This gives
\begin{equation}
\begin{split}
\left\langle \psi_{1}^{\prime}\psi_{2}^{\prime}\right\rangle& =C(\Delta y,\Delta x)+C_2(y_{+},\Delta y) \\
& +2y_{+}\sqrt{\frac{\alpha}{\epsilon}}\int_{0}^{\Delta x}dz\int_{0}^{\Delta y/2}dz'\tilde{\chi}_{12}\left(z,z'\right).
\end{split}
\end{equation}
Next, $C_2(y_{+},\Delta y)$ is in fact a zero mode of the individual advection operators $\mathcal{L}_i$, reflecting the fact that $k_x=0$ modes of $\psi'$ are not advected by a mean flow pointing in the $\hat{x}$ direction, irrespective of the shape of the mean flow (as discussed above). So, such contributions to the correlation function are not constrained by the quasi-linear approximation. While there is no a-priori reason to set them to zero, we may thus expect that modes with $k_x=0$ do not contribute significantly. Indeed, we see empirically in our DNS that when setting $\Delta y=0$ the even part of the two-point correlation function (with varying $\Delta x$) is independent of $y_+$, see Fig.~\ref{fig:2-pt_1D}(a) and Fig.~\ref{fig:2-py_f(y+)}(b). Thus, it does not contribute, at least to leading order.
Finally, the full solution reads
\begin{equation}
\begin{split}
\left\langle \psi_{1}^{\prime}\psi_{2}^{\prime}\right\rangle& =C(\Delta y,\Delta x) \\
& +2y_{+}\sqrt{\frac{\alpha}{\epsilon}}\int_{0}^{\Delta x}dz\int_{0}^{\Delta y/2}dz'\tilde{\chi}_{12}\left(z,z'\right).
\label{eq: two point sol}
\end{split}
\end{equation}

\section{Spatial and temporal resolution of DNS}
\label{appendix: resolution}
\begin{figure}[h]
\includegraphics[width=1\columnwidth]{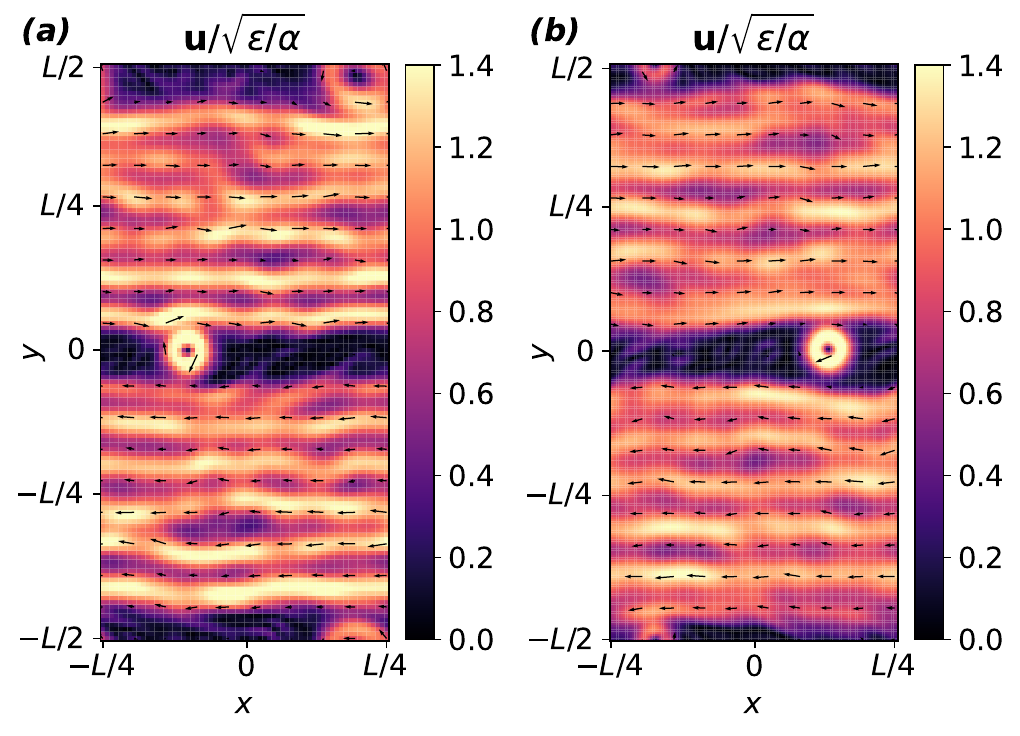}
\caption{\label{fig:hires_snap}
Comparison of the LQG jet condensate at two spatial resolutions, showing the velocity  $\boldsymbol{v}=\hat{z}\times\boldsymbol{\nabla}\psi$ snapshot (a) Simulation-B with $dx=dy=0.0491$ and (b) Simulation-B(*2) with $dx=dy=0.0245$.}
\end{figure}

\begin{figure}[h]
\includegraphics[width=1\columnwidth]{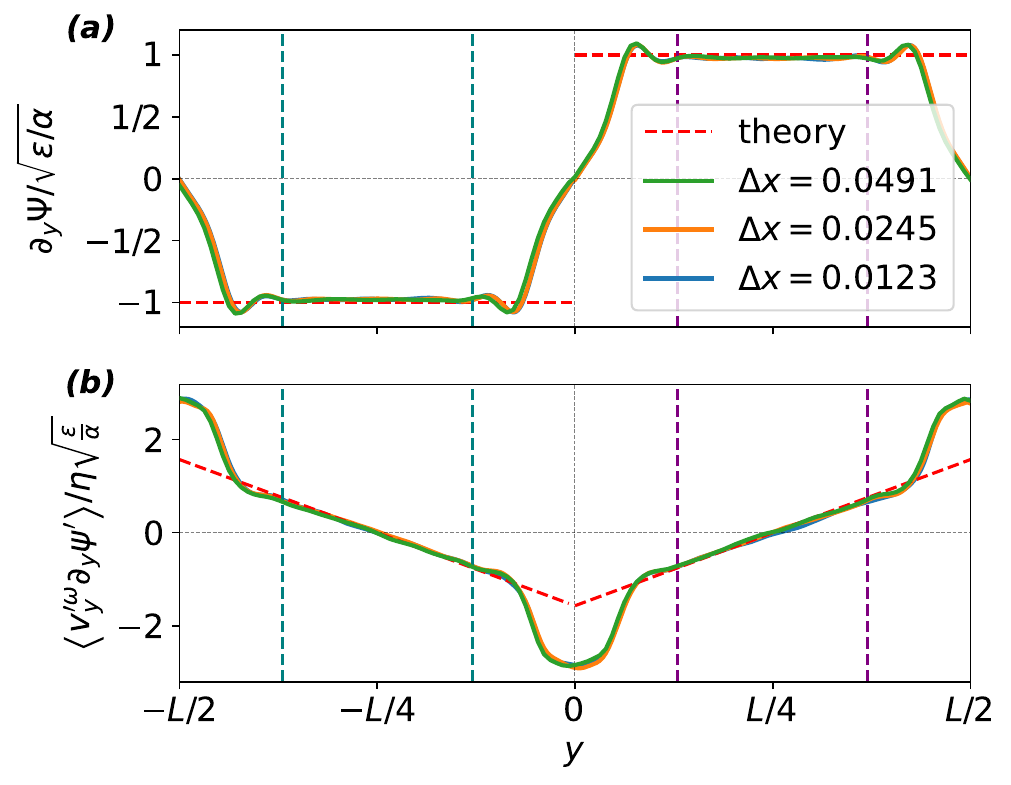}
\caption{\label{fig:hires_terms}
The terms (a) $\partial_y \Psi$ and (b) $\langle v_{y}^{\omega\prime} \partial_y \psi^{\prime}\rangle$ as measured from the DNS (solid lines), rescaled and compared with their theoretical predictions (dashed line), for simulations  B, B(*2) and B(*4) with spatial resolution $dx=dy=0.0491$, $dx=dy=0.0245$ and $dx=dy=0.0123$ respectively, with all other parameters the same. The profiles are obtained from short averaging over $\sim 30 T_L$. The two regions where the leading order solution for the mean flow applies are delimited by vertical dashed lines.}
\end{figure}

For all DNS, a constant time step $dt$, different for each DNS was used (Table~\ref{tab:timesteps}), with the forcing amplitude normalized by $\sqrt{dt}$ so that energy injection is independent of it. The grid spacing $dx=dy\approx0.05$ is the same for all simulations considered at the default resolution of 64x128. To verify the adequacy of the choice of $dt$ and $dx$, they are compared with the smallest physical time scale $\tau_E(l_\nu)$ and length scale $l_\nu$ respectively as presented in Table~\ref{tab:timesteps}. While the temporal resolution is smaller by at least 4 orders of magnitude than $\tau_E(l_\nu)$, the grid spacing is relatively close to the Kolmogorov scale $l_\nu$. The large difference between the spatial and temporal resolutions required is due to the hyper-viscosity used in the evolution equation~\eqref{eq:LQG}. It allows us to use a relatively large grid spacing (or low resolution), as for $p=7$ the energy cutoff is extremely sharp leaving only a small fraction of the total energy at length scales $l_\nu < l \leq dx$. 
To verify that the simulations are spatially fully developed and that this resolution is not too coarse, two high-resolution simulations (Sim-B(*2) and Sim-B(*4)) were performed with the same parameters as Simulation-B but with $dx = dy \approx 0.025$ and $dx = dy \approx 0.0125$ corresponding to 128x256 and 256x512 resolutions respectively. The resulting condensate is exactly the same as for the lower resolution, as can be appreciated from the snapshot comparison in Fig.~\ref{fig:hires_snap} and from the averaged terms in Fig.~\ref{fig:hires_terms}.

\begin{table}[h]
\caption{\label{tab:timesteps} Spatial and temporal resolutions used for integration of the simulations detailed in Table~\ref{tab:sim-list} with and addition of simulations $B(*2)$, $B(*4)$ and $B(p5)$ which are performed at higher spatial resolutions. The spatial resolutions $dx$ are compared with the  Kolmogorov scale $l_\nu$ while the temporal resolution $dt$ is compared with the fastest timescale $\tau_E(l_\nu)$.}

\begin{ruledtabular}
\begin{tabular}{lcccc}
\textrm{}&
\textrm{ $dx$ }&
\textrm{ $dt$ }&
\textrm{ $l_\nu / dx$ }&
\textrm{ $\tau_E(l_\nu) / dt$}\\
\colrule
A     & 0.0491 & $4.0\cdot 10^{-5}$ &  4.085  &  $5.2 \cdot 10^{3}$\\
B     & 0.0491 & $4.0\cdot 10^{-5}$ &  4.086  &  $5.2 \cdot 10^{3}$ \\
C     & 0.0491 & $2.0\cdot 10^{-5}$ &  4.087  &  $1.1 \cdot 10^{4}$ \\
D     & 0.0491 & $4.0\cdot 10^{-5}$ &  4.185  &  $7.4 \cdot 10^{3}$ \\
E     & 0.0491 & $1.6\cdot 10^{-5}$ &  3.224  &  $7.0 \cdot 10^{3}$ \\
F     & 0.0491 & $1.0\cdot 10^{-4}$ &  4.279  &  $2.9 \cdot 10^{3}$ \\
G     & 0.0491 & $4.0\cdot 10^{-5}$ &  4.278  &  $7.3 \cdot 10^{3}$ \\
H     & 0.0491 & $4.0\cdot 10^{-5}$ &  4.279  &  $7.3 \cdot 10^{3}$ \\
I     & 0.0491 & $2.0\cdot 10^{-5}$ &  4.076  &  $1.0 \cdot 10^{4}$ \\
J     & 0.0491 & $1.6\cdot 10^{-5}$ &  3.217  &  $6.7 \cdot 10^{3}$ \\
K     & 0.0491 & $1.6\cdot 10^{-5}$ &  3.216  &  $6.7 \cdot 10^{3}$ \\
B(*2) & 0.0245 & $4.0\cdot 10^{-5}$ &  8.169  &  $5.2 \cdot 10^{3}$ \\
B(*4) & 0.0123 & $4.0\cdot 10^{-5}$ &  16.34  &  $5.2 \cdot 10^{3}$ \\
B(p5) & 0.0245 & $4.0\cdot 10^{-5}$ &  8.149  &  $5.2 \cdot 10^{3}$ 
\end{tabular}
\end{ruledtabular}
\end{table}

\begin{figure}[h]
\includegraphics[width=1\columnwidth]{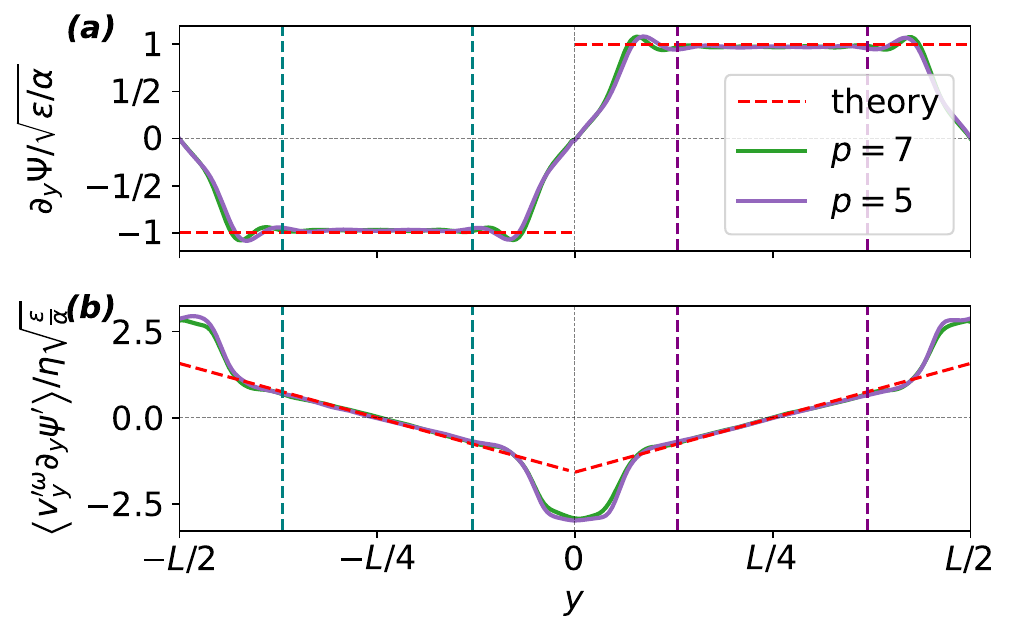}
\caption{\label{fig:hypVisc_terms}
The terms (a) $\partial_y \Psi$ and (b) $\langle v_{y}^{\omega\prime} \partial_y \psi^{\prime}\rangle$ as measured from the DNS (solid lines), rescaled and compared with their theoretical predictions (dashed line), for simulations B(*2) and B(p5) with hyper-viscocity ($\nu = 7.3\times10^{-19}$, $p=7$) and ($\nu = 6.91\times 10^{-13}$, $p=5$) respectively chosen such that $k_\nu \approx 31.2$  The two regions where the leading order solution for the mean flow applies are delimited by vertical dashed lines.}
\end{figure}

Finally, to demonstrate that the choice of hyper-viscosity in eq.~\eqref{eq:LQG} does not affect the mean flow condensate, we performed an additional low hyper-viscosity run with the same parameters as Sim-B(*2) but with $p=5$ and $\nu = 6.3\times 10^{-13}$  - Sim-B(p5). The value of $\nu$ was chosen so that the Kolmogorov scale for both Sim-B(*2) and Sim-B(p5) is $l_\nu \approx 0.2$. The higher resolution (compared to Sim-B) is used in the comparison as the energy cutoff is not as sharp in the $p=5$ case, requiring a larger separation of scales to ensure convergence. The comparison demonstrates that the choice of hyper-viscosity does not affect the mean flow, as presented in Fig.~\ref{fig:hypVisc_terms}.
Note that due to the longer time required to integrate the equations at the high resolution, only limited statistics were obtained amounting to $\sim 30 T_L$. To make the comparison quantitative, the averaged terms presented in Figs.~\ref{fig:hires_terms},\ref{fig:hypVisc_terms} are over $30 T_L$ for all simulations. 

\end{document}